\documentclass[longauth]{aa}  
\usepackage{graphicx}
\usepackage{txfonts}
\usepackage{hyperref}
\hypersetup{
    colorlinks=true,
    linkcolor=blue,
    filecolor=blue,      
    urlcolor=blue,
    citecolor=blue}
\usepackage{natbib}
\begin{document} 
   \title{Hot H$_2$O emission during an outburst in a classical T Tauri star}
   \author{Zhen Guo \inst{1, 2}\fnmsep\thanks{zhen.guo@uv.cl}
          \and         
          J. Osses\inst{1, 2}
          \and
          V. Fermiano \inst{1, 2}
          \and
          Yuting Zhou \inst{3}
          \and
          Min Fang \inst{4,5}
          \and
          G. Herczeg \inst{6, 7}
          \and
          A. Carvalho \inst{8, 9}
          \and
          V. Elbakyan\inst{10,11}
          \and
          L. Hillenbrand  \inst{9}
          \and
          Mutian Wang \inst{12}
          \and
          Hanpu Liu \inst{13}
          \and
          Yao Liu  \inst{14, 4}
          \and
          C.~Briceño\inst{15}  
          \and
          K. Singh\inst{16, 17} 
          \and
          V. D. Ivanov\inst{18}        
          \and
          J. Ninan\inst{17}           
          \and  
          T. Giannini \inst{19}
          \and
          A. Aliaga\inst{1,20} 
          \and
          C. Morris\inst{1} 
          \and
          M. Montesinos\inst{21}
          \and
          He Zhao\inst{22, 23}
          \and
          C.~Contreras~Peña\inst{24, 25} 
          \and
          J. Jose \inst{26} 
          \and
          T. Chand \inst{27} 
          \and
          Wen-Ping Chen \inst{28} 
          \and
          Wei-Hao Wang \inst{29} 
          \and
          Yang Huang \inst{30, 31} 
          \and
          C. Lopez\inst{1} 
          \and
          P. Fernandez-Schlosser\inst{1} 
          \and
          D. Correa-Herrera\inst{1} 
          \and  
          R. Kurtev\inst{1}
          \and
          V. Rodriguez\inst{1} 
          \and
          C. Lizana-Vidal\inst{1} 
          \and
          J. Borissova\inst{1} 
          \and 
          M. Kuhn\inst{16} 
          \and 
          R. K. Saito\inst{32} 
          \and 
          Ram K. Yadav\inst{33}
          }
   \institute{Instituto de F{\'i}sica y Astronom{\'i}a, Universidad de Valpara{\'i}so, ave. Gran Breta{\~n}a, 1111, Casilla 5030, Valpara{\'i}so, Chile    \\
              \email{zhen.guo@uv.cl}
         \and
Chinese Academy of Sciences South America Center for Astronomy (CASSACA), National Astronomical Observatories, CAS,
Beijing 100101, China
    \and
High School Affiliated to Nanjing Normal University, Nanjing, China
       \and
       Purple Mountain Observatory, Chinese Academy of Sciences, 10 Yuanhua Road, Nanjing 210023, People’s Republic of China; 
        \and
        University of Science and Technology of China, Hefei 230026, People’s Republic of China
        \and
        Kavli Institute for Astronomy and Astrophysics, Peking University, No. 5 Yiheyuan Road, Haidian District, Beijing 100871, People’s Republic of China
        \and
         Department of Astronomy, Peking University, No. 5 Yiheyuan Road, Haidian District, Beijing 100871, People’s Republic of China
         \and
         Center for Astrophysics — Harvard \& Smithsonian, Cambridge, MA 02138, USA
         \and
         Department of Astronomy; California Institute of Technology; Pasadena, CA 91125, USA
        \and
           Fakultat fur Physik, Universitat Duisburg-Essen, Lotharstraße 1, D-47057 Duisburg, Germany
        \and
        Research Institute of Physics, Southern Federal University, Rostov-on-Don 344090, Russia
        \and
        School of Astronomy and Space Science, Nanjing University, Nanjing 210023, China
         \and
         Department of Astrophysical Sciences, Princeton University, 4 Ivy Lane, Princeton, NJ 08544, USA
          \and
          School of Physical Science and Technology, Southwest Jiaotong University, Chengdu 610031, China
           \and
          Cerro Tololo Inter-American Observatory, National Optical Astronomical Observatory, Casilla 603, La Serena, Chile
           \and
           Centre for Astrophysics Research, University of Hertfordshire, Hatfield AL10 9AB, UK
        \and
         Department of Astronomy and Astrophysics, Tata Institute of Fundamental Research, Mumbai, 400005, India
         \and
        European Southern Observatory, Karl-Schwarzschild-Strasse 2, D-85748 Garching bei München, Germany
         \and
         INAF—Osservatorio Astronomico di Roma, Via di Frascati, 33, 00078, Monte Porzio Catone, Italy
                 \and
        Millennium Nucleus on Transversal Research and Technology to Explore Supermassive Black Holes (TITANS)
            \and
          Departamento de Física, Universidad Técnica Federico Santa María, Avenida España 1680, Valparaíso, Chile
        \and 
       Institute of Astronomy and Physics, Inner Mongolia University, Hohhot 010021, People's Republic of China
       \and 
       Departamento de Fisica y Astronomia, Facultad de Ciencias Exactas, Universidad Andres Bello, Fernandez Concha 700, 8320000 Santiago, Chile
        \and 
        Department of Physics and Astronomy, Seoul National University, 1 Gwanak-ro, Gwanak-gu, Seoul 08826, Republic of Korea
        \and
       Research Institute of Basic Sciences, Seoul National University, Seoul 08826, Republic of Korea
       \and
       Department of Physics, Indian Institute of Science Education and Research Tirupati, Yerpedu, Tirupati - 517619, Andhra Pradesh, India
       \and
         Aryabhatta Research Institute of Observational Sciences (ARIES), Manora Peak, Nainital, 263001, India 
       \and
        Institute of Astronomy, National Central University, 300 Zhongda Road, Zhongli 320317 Taoyuan, Taiwan
        \and
        Institute of Astronomy and Astrophysics, Academia Sinica, Taipei 10617, Taiwan
         \and
       School of Astronomy and Space Science, University of Chinese Academy of Science, Beijing 100049, China
       \and
       National Astronomical Observatories, Chinese Academy of Sciences, Beijing, 100101, China
       \and
        Departamento de Física, Universidade Federal de Santa Catarina, Trindade 88040-900, Florianópolis, Brazil
        \and
        National Astronomical Research Institute of Thailand Sirindhorn AstroPark, 260 Moo 4, T. Donkaew, A. Maerim, Chiangmai, 50180 Thailand
       }

 \date{Received xxx; accepted xxx}
   \abstract
  {The unstable mass accretion process in young stellar objects (YSOs) often triggers observable outbursts. These episodic accretion events play a critical role in stellar mass assembly during the pre-main-sequence phase. In this paper, we present observations of an eruptive young star in the Rosette Nebula, identified by the \textit{Gaia} Science Alerts system using \textit{Gaia} time-series data.}
   {We aim to investigate the evolution of the brightness and mass accretion rate of V557~Mon throughout its outburst and subsequent decline. In addition, we trace the evolution of the inner accretion disk during the outburst by monitoring molecular emission features.}
   {We compiled multi-band photometric time-series from Gaia, ZTF, and several 1 m-class ground-based telescopes and obtained optical and near-infrared spectra at multiple epochs covering the outburst and fading phases. Stellar parameters were derived from quiescent colour, spectra and spectral energy distribution (SED) fitting. We also measured the mass accretion rate and fit models to molecular emission bands.}
   {Since late 2024, V557~Mon has undergone a year-long outburst consistent with EXor variability. Based on quiescent photometry, V557~Mon has a spectral type of M1 with an extinction of $A_V = 1.8\pm0.3$~mag, consistent with a 0.4--0.5~$\rm M_\odot$ star at an age of 2~Myr. Our multi-epoch spectra and $u$-band photometry indicate a peak accretion rate of $6.3\times10^{-7}$~$\rm M_\odot\,yr^{-1}$ during the outburst, roughly 70 times higher than in quiescence. We report the detection of hot water vapour emission bands, together with TiO, VO, and CO emission features. Using ExoMol models, we measured that the inner-disk temperature changed from 3000~K to 2000~K during the fading phase of the outburst.}
   {We report a recent EXor outburst in a low-mass Class~II YSO. Our observations reveal the transient formation of a hot molecular inner disk, traced by variable water vapour emission during the EXor event. A positive correlation is found between the molecular excitation temperature and the overall stellar brightness.}
   \keywords{Stars: variables: T Tauri, Herbig Ae/Be;  Stars: pre-main sequence; Infrared: stars}
 
\maketitle

\section{Introduction}
During the pre-main-sequence evolutionary stage, young stars accrete mass from their circumstellar disks. In some cases, the mass accretion rate increases by several orders of magnitude, producing optical and infrared outbursts detectable through time-domain surveys \citep[see reviews by][]{Audard2014, Fischer2023}. These episodic accretion events are thought to play a critical role in building stellar mass and may help resolve the luminosity problem \citep[e.g.][]{Hartmann1998, Dunham2012}. Such bursts release heat and energy into the disk, shifting snow lines and driving chemical evolution \citep{Cieza2016, Kospal2023}, thereby influencing the formation of planetesimals \citep{Ros2024}.

Two principal families of eruptive variables are commonly distinguished among young stars based on their observational characteristics. EXor objects, first exemplified by EX~Lupi, undergo optical brightening events of a few magnitudes lasting several hundred days, whereas FUor-type objects display higher-amplitude outbursts that can persist for decades \citep{Herbig1977eruptive, Hartmann1996}. {The year-long duration accretion outburst, so-called EXors, is thought to be a common stage during the evolution of Class II YSO. In total, about 16 EXors have been identified by photometric surveys \citep[see a summary of these objects in][]{Contreras2025}, with a typical duration of 1 year. On average, 1-2 EXors are identified annually in recent years.} We have yet to find a regular pattern of these eruptive events, although some EXors show repeated outbursts \citep{Singh2026, Matefy2026}. 

These observational outbursts trace episodes of enhanced disk accretion but differ in amplitude, duration, and spectral characteristics, reflecting distinct regimes of interaction among the star, magnetic field, and inner disk. For instance, \citet{Guo2024a} found that FUor-type events show higher mid-infrared amplitudes than those on EXors due to extensive heating of the circumstellar disk. Some objects exhibit photometric or spectroscopic behaviour intermediate between EXor and FUor outbursts \citep[e.g.][]{Briceno2004, Guo2025}, indicating more complex physical mechanisms behind the observation. 

Several theories have been proposed to explain the mechanisms driving year-long accretion outbursts. These include instabilities at the magnetospheric boundary \citep{DAngelo2010, Armitage2016}, thermal instabilities in the inner disk \citep{Nayakshin2024}, tidal perturbations \citep{Montesinos2026}, or perturbations produced by a planetary mass object on an eccentric orbit \citep{Dunhill2015, Teyssandier2020, YHLee2020}. These short-lived outbursts differ from the decades-long FUor-type events, which may be triggered by disk fragmentation \citep{Vorobyov2010}, dynamical interactions \citep{Lodato2004, Cuello2019, Nayakshin2024b}, or by a combination of magneto-rotational and gravitational instabilities \citep{Zhu2009}.

During EXor outbursts, the mass accretion rate can increase by up to three orders of magnitude \citep{Cruz-Saenz2023, Wang2023}, while still being regulated by the stellar magnetic field \citep{Lorenzetti2012}. Photometrically, these events are marked by brightening in the optical and near-infrared (NIR) bands, and spectroscopically by strong emission lines, including the hydrogen recombination series, helium, and metallic lines, which trace magnetospheric accretion shocks and the heated inner disk regions \citep[e.g.][]{Lorenzetti2009}. Systematic optical and infrared monitoring campaigns have recently uncovered numerous new eruptive systems, revealing a broad diversity in their temporal and spectral behaviour \citep[e.g.][]{Guo2021, Siwak2023, Kuhn2024, Nagy2025, Giannini2026}. A recent catalogue summarises both confirmed and candidate eruptive events in YSOs \citep{Contreras2025}.

In late 2024, an optical outburst was discovered on a young stellar object (YSO) V557~Mon (06:33:31.31 +04:52:37.2, J2000) located in the Rosette Nebula, a prominent H~{\sc ii} region hosting the young open cluster NGC~2244 \citep[][]{Carvalho2025}. The cluster NGC~2244 has an estimated age of about 2~Myr and is centred at a distance of 1.44$\pm0.03$~kpc \citep{Muzic2022}. Based on the infrared spectral energy distribution (SED), V557~Mon was classified as a Class~II YSO, showing infrared excess consistent with an actively accreting disk \citep{Kuhn2021}. V557~Mon was classified as a variable star first in \citet{Kukarkin1970}. During the 2024 outburst, this source brightened by more than 3.5~mag in the \textit{Gaia} $G$ band between August and November 2024. This event was independently classified as a variable YSO by the ALeRCE broker \citep{Alerce2021}, based on Zwicky Transient Facility (ZTF) time-series data. Near-infrared spectroscopy obtained in January 2025 with the Palomar 200‑inch telescope revealed a spectrum dominated by strong atomic and molecular emission lines, including H, He, and CO bandheads, as well as a continuum excess originating from a hot, bright inner disk \citep[][]{Carvalho2025}. These features are consistent with an EXor outburst commonly observed in Class~II objects. 

In this paper, we present a comprehensive multi-wavelength photometric and spectroscopic study of the 2024--2025 outburst of V557~Mon. We first determined the stellar parameters using quiescent spectra and photometry, then measured the event duration from time-series data. We measured the mass accretion rate from UV-excess photometry and optical-to-NIR emission lines. Finally, we analysed molecular emission features (CO and H$_2$O), providing the first time-series observations of NIR H$_2$O emission bands during an EXor outburst. This is the first time that broadband CO, H2O, and TiO/VO are observed in emission around a classical T Tauri star (CTTS), making V557~Mon a unique target for studying the inner disk evolution during an outburst.

\begin{figure}
  \centering
     \includegraphics[height=5.5cm]{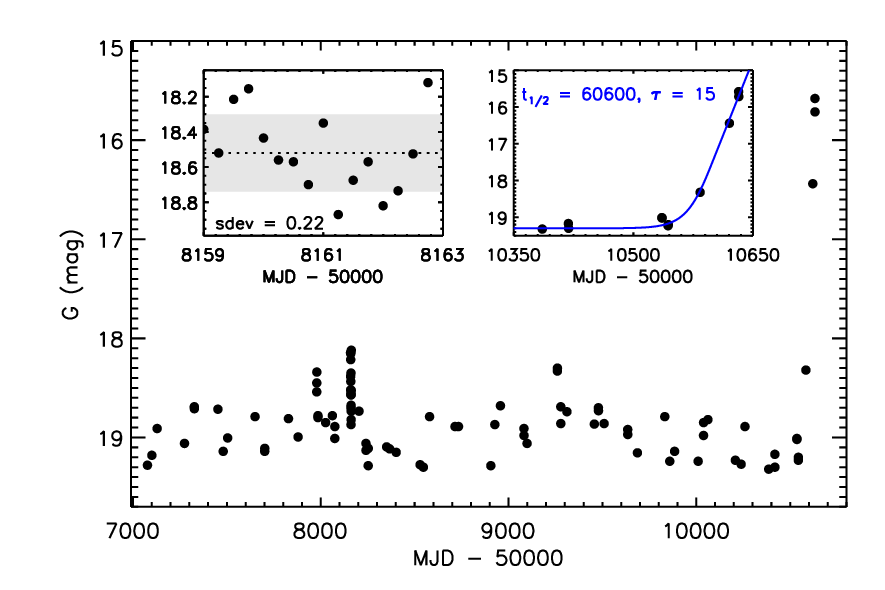}
     \caption{\textit{Gaia} alert $G$-band light curve of V557~Mon. \textit{Upper left}: High-cadence time-series taken in December 2018. The mean brightness (dotted line) and standard deviation (shaded grey area) are shown. \textit{Upper right}: Rising curve of the outburst in 2024. A two-step exponential-linear function is presented in blue with $t_{1/2} = 60600$ d and $\tau = 15$ d. }
\label{fig:Gaia_lc}
\end{figure}
\section{Observational data}

\subsection{Photometric data}
In this section, we present the photometric data of V557~Mon during its quiescent and the recent eruptive stage.

\subsubsection{Gaia science alert}
The 2024 outburst of V557~Mon was identified as a $\sim$3.5~mag brightening event in a known YSO, catalogued as Gaia24djk \citep{Hodgkin2021}. We retrieved its $G$-band light curve from the \textit{Gaia} Science Alerts database (see Fig.~\ref{fig:Gaia_lc}). Prior to the outburst, V557~Mon exhibits a mean brightness of $G=18.95$~mag with a standard deviation of $\delta G=0.26$~mag, attributable to short-term variability (see the inset panel in Fig.~\ref{fig:Gaia_lc}). V557~Mon has a geometric distance of $1.67^{+0.67}_{-0.51}$~kpc based on the \textit{Gaia} EDR3 \citep{Bailer-Jones2021}. The distance to V557~Mon is consistent with the mean distance of NGC~2244 members \citep[][]{Muzic2022}. Therefore, in the following analysis, we adopt a distance of $1.44\pm0.03$~kpc for V557~Mon to avoid the large error bar from the individual parallax measurement. 

\begin{figure*}
  \centering
     \includegraphics[height=8.3cm]{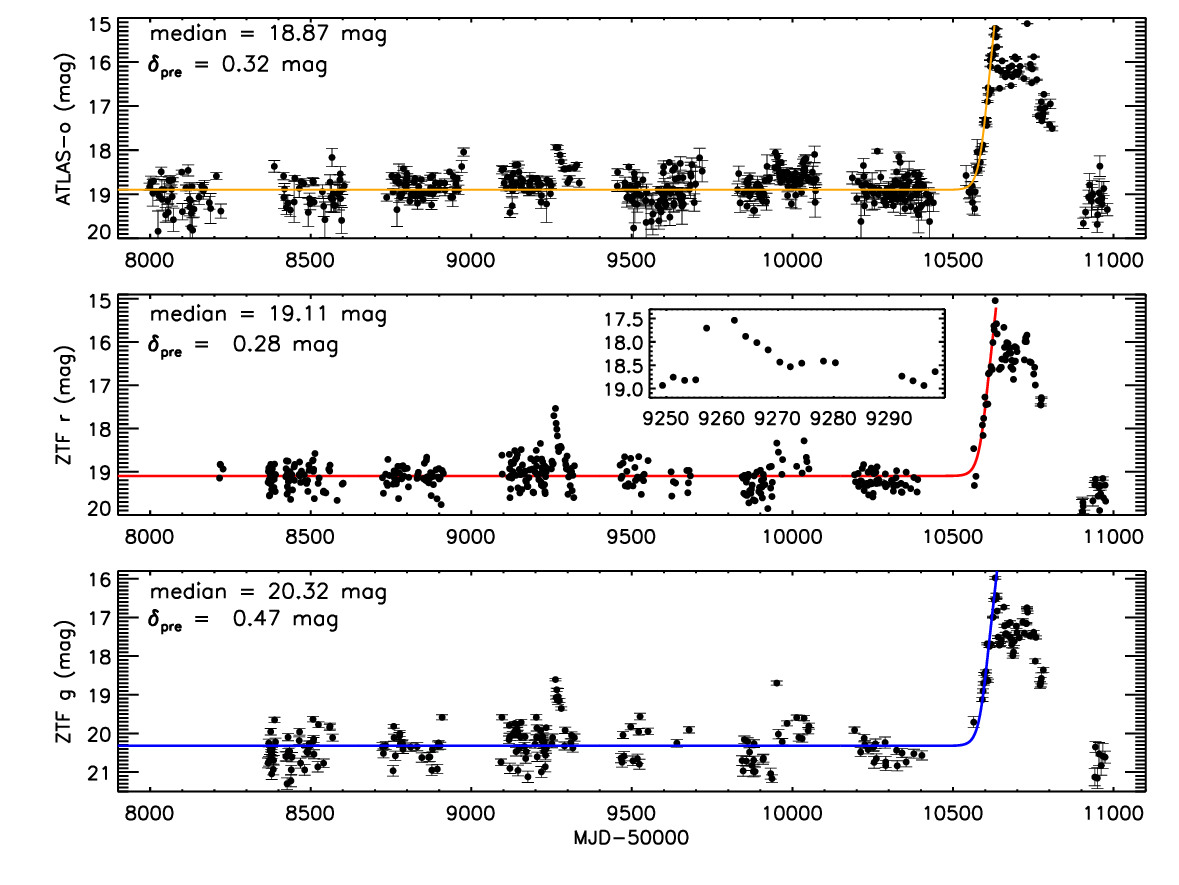}
          \includegraphics[height=8.5cm]{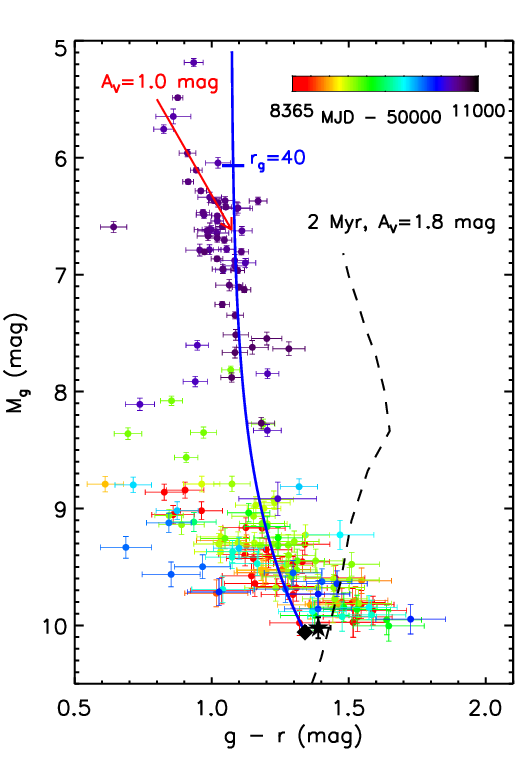}
     \caption{\textit{Left:} Multi-band time-series of V557~Mon from October 2015 to October 2025. Rising curves are fitted to each light curve. The median brightness and the standard deviation ($\delta_{pre}$) during the pre-outburst stage (MJD $<$ 60500) are written on the plot. A subplot in the middle panel presents a previous short-timescale burst. \textit{Right:} Colour-magnitude diagram of V557~Mon. Data points were colour-coded by their observation time. We present the 2~Myr isochrone of the BHAC15 evolution track, reddened by $A_V = 1.8$~mag and $R_V = 3.1$ \citep[using extinction law from][]{WangS2019}. We also show a veiling curve up to $r_g = 40$ \citep[blue line;][]{Herczeg2025}. The black diamond and star represent photometry obtained from SOAR and Pan-STARRS, respectively. The Pan-STARRS photometry is converted to the SDSS filters.}
\label{fig: ztf_all}
\end{figure*}
\subsubsection{ZTF and ATLAS photometry}

Several ground‑based time‑domain surveys monitored the brightness variations of V557~Mon, including ZTF~\citep[][]{Bellm2019} and ATLAS \citep[Asteroid Terrestrial-impact Last Alert System,][]{Tonry2018}. We retrieved the latest ZTF $g$ and $r$-band light curves from the ALeRCE ZTF Explorer\footnote{Automatic Learning for the Rapid Classification of Events \citep{Alerce2021}. \url{https://alerce.online}}. We obtained the ATLAS $o$-band time-series through the Black Hole Target Observation Manager (BHTOM)\footnote{\url{https://bh-tom2.astrolabs.pl/}}, where V557~Mon is listed under the identifier Gaia24djk~\citep{BHTOM2025}. For the ATLAS data, we combined the epoch photometry into 1-day bins to increase the signal-to-noise ratio (S/N).

\subsubsection{Multi-band photometric monitoring}
We obtained multi-band photometry of V557~Mon to supplement the ZTF and ATLAS survey light curves. The primary observations were acquired on the 1 m telescopes at the Las Cumbres Observatory Global Telescope Network \citep[LCOGT;][]{Brown2013} using the $u$, $g$, $r$, and $i$ filters. The pixel size of the CCD is 0.4\arcsec\. Bias-subtracted and flat-field–corrected science frames were retrieved from the LCO public archive. We built a data reduction pipeline in Python. For each night and filter, individual frames were co‑added and visually inspected to verify alignment. Aperture photometry was carried out on the co‑added images using a fixed 6‑pixel‑radius aperture centred on the target. The local sky background was estimated from an annulus with inner and outer radii of 10 and 15 pixels, respectively. The mean sky background within each annulus was measured using sigma‑clipped statistics to exclude outliers and subtracted from the raw aperture fluxes. Photometric uncertainties were computed following standard Poisson statistics, including contributions from object counts, background counts, and readout noise. We assessed several nearby comparison stars for variability and adopted non‑variable sources as photometric references. Relative magnitudes of V557~Mon in each band were calibrated against the catalogued magnitudes of these reference stars from SkyMapper \citep{Onken2024}, with photometric uncertainties less than 0.05~mag for each reference star.

From January 11 to February 10 2025, we took multi-epoch $V$, $R$, and $I$-band photometry of V557~Mon using the 40~cm telescope at Lulin Observatory. We obtained photometric data from the 1~m SWOPE telescope at Las Campanas Observatory, in the SDSS $u$, $g$, $r$, and $i$ bands. The observations were taken daily between December 27 and January 7. We co-added the $u$-band images over a time window of a few days to reach a higher S/N. We also obtained $U$, $B$, $V$, $R$, and $I$ band photometric observations with two 70 cm Thai Robotic Telescopes (TRT-CTIO and TRT-SBO), operated by the National Astronomical Research Institute of Thailand. We conducted WCS calibration using the TRT pipeline and online tools\footnote{\url{astrometry.net}}. The photometric data are presented in the appendix.

\begin{figure*}
  \centering
     \includegraphics[height=5.4cm]{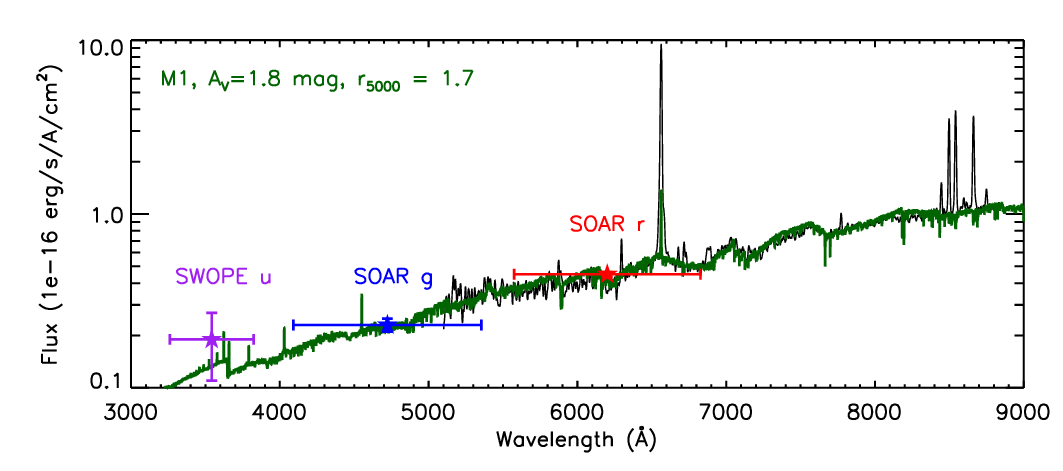}
      \includegraphics[height=5.1cm]{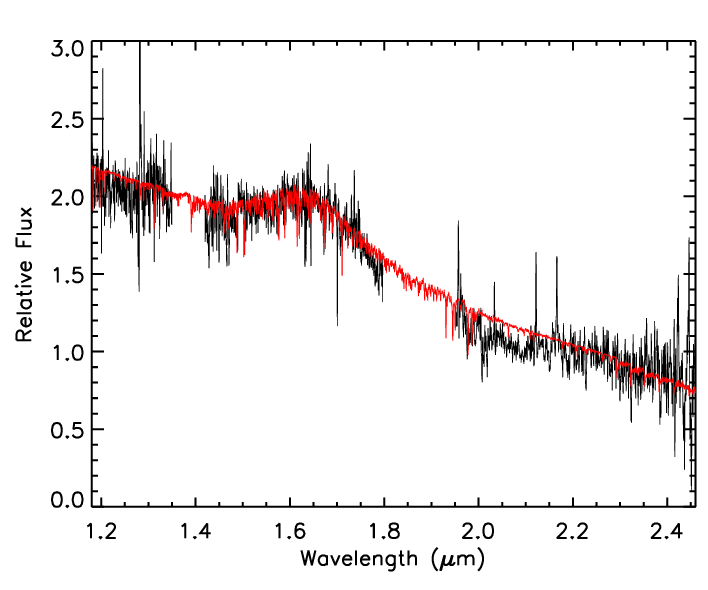}
     \caption{Optical (left) and NIR (right) spectra of V557~Mon after the 2024--2025 outburst. A best-fit M1-type photospheric template is overplotted in green, with $A_V = 1.8$~mag and $r_{5000} = 1.7$. The photometric data ($u$, $g$, $r$, $i$) were observed by SOAR and SWOPE. A toy model constructed from a 3600~K photosphere and a 1500~K blackbody is overplotted in red.}
\label{fig: qspec}
\end{figure*}

\subsection{Spectroscopic data}

The spectroscopic follow-up observation of V557~Mon was conducted under the `caught-on-fire' campaign with observation time allocated by the Chilean Telescope Allocation Committee on the Magellan and SOAR telescopes.

\subsubsection{Magellan FIRE spectra}

We observed two NIR spectra of V557~Mon in the outburst (January 10 2025) and the fading stage (April 29 2025), using the FIRE spectrograph \citep{Simcoe2013} on the Magellan Baade Telescope. We utilised the 0.6$"$ slit (R = 6000) with a wavelength range between 0.8 and 2.5~$\mu$m and a spatial resolution of 0.18 arcsec per pixel. The exposure time is $240\rm s\times4$ in January and $ 300\rm s\times4$ in April. We used telluric standard stars (A0V) observed immediately after the scientific exposures for telluric correction and flux calibration. We reduced the data using the FireHose pipeline \citep{gagne2015} and conducted telluric correction using \textit{xtellcorr} \citep{Vacca2003}. We performed the following customised calibration routines for the FIRE data. First, we added a third-order polynomial wavelength solution using OH skylines. Second, we fitted a linear slope at the edge among the last three spectral orders in K-bandpass, to remove the curvature on the continuum introduced by the FireHose pipeline \citep[see examples in][]{Guo2020}. After the telluric correction by the standard star, we performed absolute flux calibration using the photometric magnitude from the $J$-band acquisition image, to compensate for the slit loss and atmospheric dispersion through the non-parallactic angle. The pipeline-reduced spectrum was integrated over the $J$-band response curve, and the total flux was scaled to match the photometric flux derived from the acquisition image.

\subsubsection{SOAR TripleSpec4.1 spectra}
We obtained multi-epoch NIR spectra using the TripleSpec4.1 (TSpec) spectrograph \citep{Schlawin2014} on the SOAR telescope at the Cerro Pachon Observatory. We chose the 1\arcsec\ slit, providing a spectroscopic resolution of $R=3500$ spanning 0.94 -- 2.46 $\rm \mu$m. The observations were conducted in AEON (Astronomical Event Observatory Network) queue mode using the standard ABBA nodding scheme, with a single exposure time ranging from 120s to 300s (based on the current brightness of the star). An A0V-type standard star was observed immediately after each science epoch for telluric correction and flux calibration. We reduced the TSpec data using the IDL-based reduction pipeline. Additionally, we included in our analysis the NIR spectrum obtained by the TSpec spectrograph on the P200 telescope, observed on January 16 2025. This spectrum was originally published in \citet{Carvalho2025}, in which the authors identified V557~Mon as an EXor outburst based on the abundance of emission features. 

\subsubsection{SOAR Goodman spectra}

We obtained three epochs of optical spectra of V557~Mon from the Goodman spectrograph mounted on the SOAR telescope, including one epoch months after the outburst. We used the low-resolution 400M2 grism with a 1\arcsec\ slit and 2x2~binning, delivering a resolving power of 13~\AA$\,$ with 3-pixel sampling and a spectral coverage between 5000 and 9050~\AA. The 1D spectroscopic data were extracted and wavelength-calibrated by the Goodman pipeline. We performed flux calibration using the instrumental curve derived from a standard star observed on the same night and the multi-band ($g$, $r$, and $i$) images obtained immediately after the spectroscopic observation.

\section{The quiescent stage}

\subsection{The pre-outburst time-series}
The pre-outburst light curves are presented in Fig.~\ref{fig:Gaia_lc} and the left panel of Fig.~\ref{fig: ztf_all}, where V557~Mon displayed short-timescale ($\sim$days) variability in optical light curves. We provide some zoomed-in light curves on the $G$ and $r$-band time-series. From the \textit{Gaia} light curve, we found that V557~Mon exhibited a variability amplitude of 0.8~mag over a timescale of only two days.

To quantify the short-timescale variability, we defined $\delta_{\rm pre}$ as the standard deviation of brightness before the 2024 outburst (until MJD 60500). We found bluer bands have larger $\delta_{\rm pre}$, as $\delta_{\rm pre} = 0.28$ in the $r$ band, $\delta_{\rm pre} = 0.32$ in the $o$ band, and $\delta_{\rm pre} = 0.47$ in the $g$ band. The ratio between $g$- and $r$-band variation is slightly deeper than the extinction vector assuming $R_V = 3.1$ \citep[$A_g/A_r = 1.42$;][]{WangS2019}. We performed periodogram analyses on the time-series data and found no robust periodicity, indicating a stochastic nature of the short-timescale variation. In the following analysis, we assume that the natural spread of the optical light curve is of the order of $\delta_{\rm pre}$ (0.3 -- 0.5 mag), greater than the typical photometric error (0.1 mag).  We also discovered a small burst in 2021 ($\Delta r = 1.5$ mag) lasting 15 days, which has been observed by NEOWISE with an amplitude of 
$\Delta W2 = 0.6$~mag. 

We present a $g$ versus $g-r$ colour-magnitude diagram (CMD) in the right panel of Fig.~\ref{fig: ztf_all}, assuming a distance of 1.44 kpc. The majority of the data points are from ZTF and LCOGT, colour-coded by the observation date. We present the photometry from Pan-STARRS DR1 (before the outburst)\footnote{The Pan-STARRS data are converted to SDSS filters, using functions provided by \citep{Tonry2012}.} and SOAR (after the outburst), which have a high S/N. The pre-outburst ZTF photometry has a large spread on the CMD, partially due to the large $g$-band photometric error. Despite the error bars, we observe a bluer-when-brighter trend, likely caused by either stochastic accretion or variable extinction. This routine variability is commonly observed on Class II YSOs. Additionally, we added a track illustrating the $g$-band veiling during the accretion burst, calculated from a hydrogen slab model \citep{Valenti1993, Herczeg2025}.

\begin{figure}[!t]
  \centering
     \includegraphics[height=6cm]{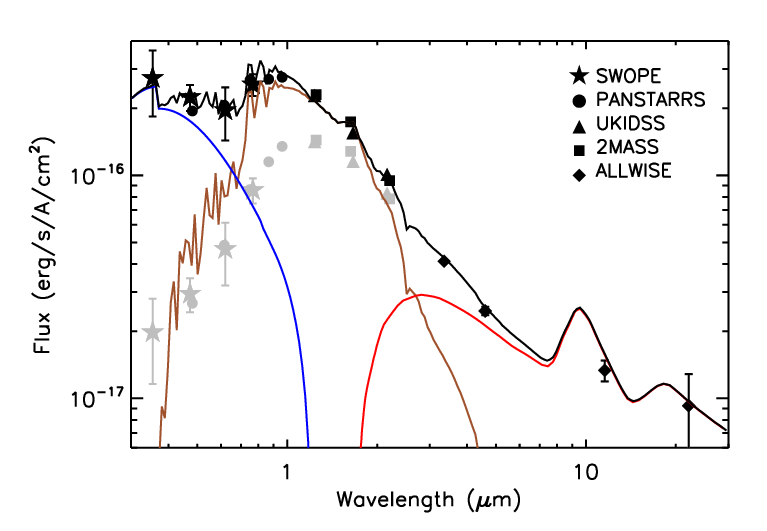}
     \caption{Quiescent SED of V557~Mon. Photometric data were obtained from Pan-STARRS, UKIDSS, 2MASS, and ALLWISE surveys (grey symbols). We included the SWOPE $u$, $g$, $r$, and $i$ band data taken after the outburst. We dereddened the SED by $A_V = 1.8$~mag (black symbols). Three components were fitted to the SED: a hot hydrogen slab (blue), a BT-Settl photospheric model (brown), and an accretion disk model from RADMC-3D \citep[red,][]{Dullemond2012}.}
\label{fig: SED}
\end{figure}

\begin{figure}
  \centering
     \includegraphics[height=4.0cm]{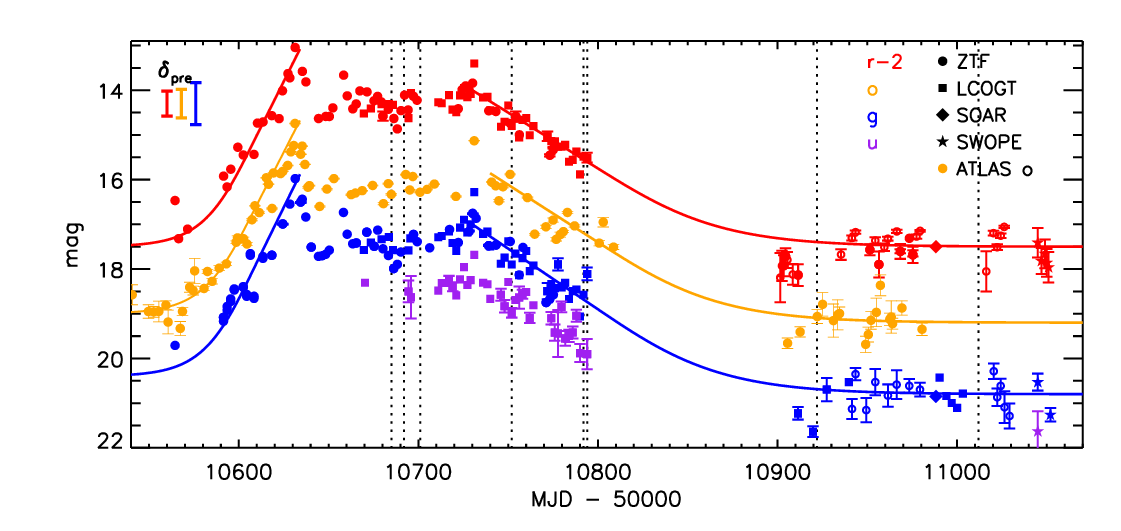}
       \includegraphics[height=4.cm]{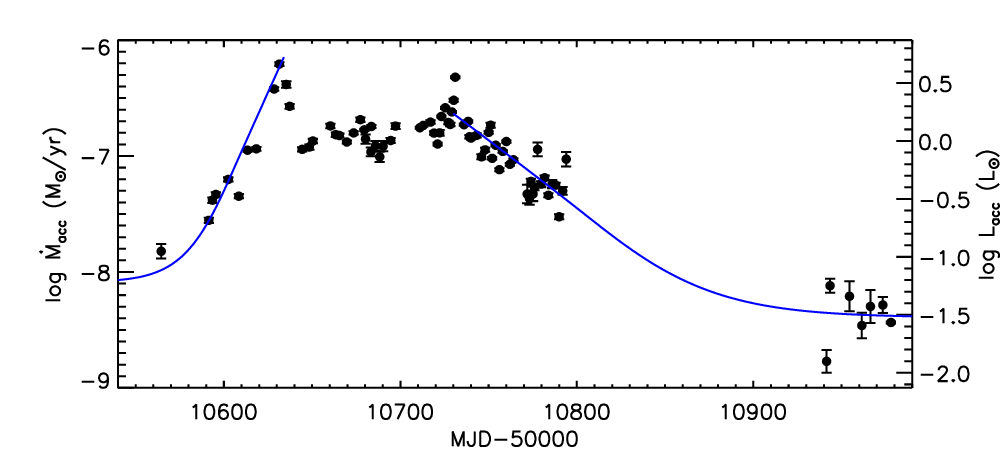}
     \caption{\textit{Upper:} Optical light curves of V557~Mon. The ZTF data includes detections (dots) and forced photometry (open circles). The pre-outburst standard deviation of $\delta_{pre} = 0.32$~mag is presented. We fit rising and fading curves to the time-series. Spectroscopic epochs are marked as vertical dashed lines. \textit{Lower:} Measured mass accretion rate ($\dot{M}_{\rm acc}$) and mass accretion luminosity ($L_{\rm acc}$) throughout the outburst.}
\label{fig: outburst_lc}
\end{figure}

\subsection{The quiescent spectra and SED}
\label{sec: quiescent}

Here, we first used the post-outburst optical spectrum obtained on December 13 2025 to estimate the spectral type (SpT), line-of-sight extinction (A$_V$), and veiling ($r_{5000}$) by an optical spectroscopic fit.  We compared our observed spectrum with the photospheric templates of weak-line T Tauri stars from \citet{Claes2024}, plus a blue excess from an accretion shock \citep[hydrogen slab model from][$T = 9000$~K]{Valenti1993}. The photospheric templates have SpT as M0, M0.5, M1, and M2. For each template, we used $A_V$ and $r_{5000}$ as free parameters (grid size of 0.1~mag and 0.1) and added extinction and veiling effects to the photospheric templates. We resampled the wavelength axis of the template to match the resolution of the Goodman spectrum. Then, we compared spectral models with our observed spectrum, applying the minimum $\chi^2$ method. After visual inspection, our best-fit result is an M1-type YSO with $A_V = 1.80$~mag and $r_{\rm 5000} = 1.7$ (see Fig.~\ref{fig: qspec}). The estimation of stellar parameters has uncertainty due to the limitations of our current dataset, and it relies on the photospheric templates and the hydrogen slab model. As discussed in \citet{Herczeg2025}, degeneracies among $A_V$, SpT, and veiling become significant with low-S/N $u$-band photometry or a lack of Balmer continuum spectra. Typically, both decreasing the line-of-sight extinction and increasing the veiling will result in a bluer spectrum. One can also increase the veiling on a late-type template to enhance the blue-band excess. 
Thus, we assumed an uncertainty of $\pm$0.3~mag in extinction, with the SpT estimated between M0 ($A_V$= 1.8~mag, $r_{5000} = 1.0$) to M2 ($A_V$= 2.0~mag, $r_{5000} = 3.0$). We present some other fitting results in the appendix. Adopting $A_V$= 1.8~mag and d = 1.44 kpc, we calculated that the H~$\alpha$ luminosity at the quiescent stage is 0.003 L$_\odot$.  

A low-S/N NIR spectrum was observed when the optical brightness of V557~Mon had faded to the pre-outburst level (September 2025). The triangular-shaped $H$-band continuum is consistent with a late-type star. We dereddened the spectrum by $A_V = 1.8$~mag, and we fitted the dereddened data with a 3600~K BT-Settl model plus a 1500~K blackbody (representing the inner accretion disk). Assuming $d=1.44$~kpc, the integrated flux of the blackbody emission is 0.035~L$_{\odot}$, about 10\% of the bolometric luminosity, higher than a typical disk around a low-mass Class II YSO. This is understandable as the spectrum was taken during the decaying stage of the outburst.  Adopting the SpT--$T_{\rm eff}$ relationship from \citet{Herczeg2014}, this temperature corresponds to an M1.8 type. The late NIR SpT has been previously reported as a result of cool spot activity \citep{Gully2017, Perez-Paolino2025}, which is common among low-mass young stars \citep{Guo2018b}.

The pre-outburst SED of V557~Mon combines photometry from the Pan-STARRS, UKIDSS, 2MASS, and ALLWISE surveys, although they were not observed simultaneously. For a stochastic variable star, we only use these data points as an example of the quiescent SED. Fig.~\ref{fig: SED} shows the dereddened SED with $A_V = 1.8$~mag. We modelled the SED with three components: (1) a stellar photosphere, (2) an accretion excess, and (3) a circumstellar disk. The photospheric contribution was represented by a BT-Settl model with $T_{\rm eff} = 3600$~K, $\log g = 3.0$, and solar metallicity \citep{Allard2012}. The accretion excess was approximated by a hydrogen slab model.

\begin{figure}
  \centering
     \includegraphics[height=6cm]{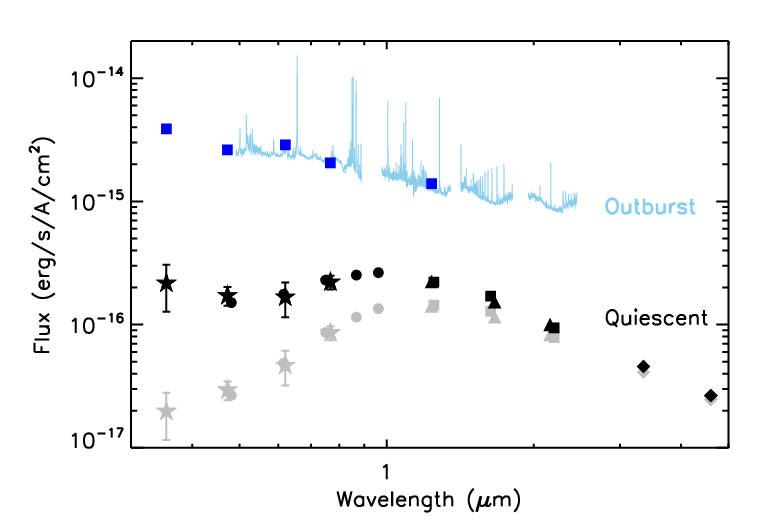}
     \caption{Dereddened SED observed during the 2024--2025 outburst (blue dots and spectra). The quiescence SED is also presented as a reference (grey: original; black: dereddened), with the same symbols in Fig.~\ref{fig: SED}.}
\label{fig:outburst_SED}
\end{figure}
\begin{figure*}
  \centering
     \includegraphics[height=7.cm]{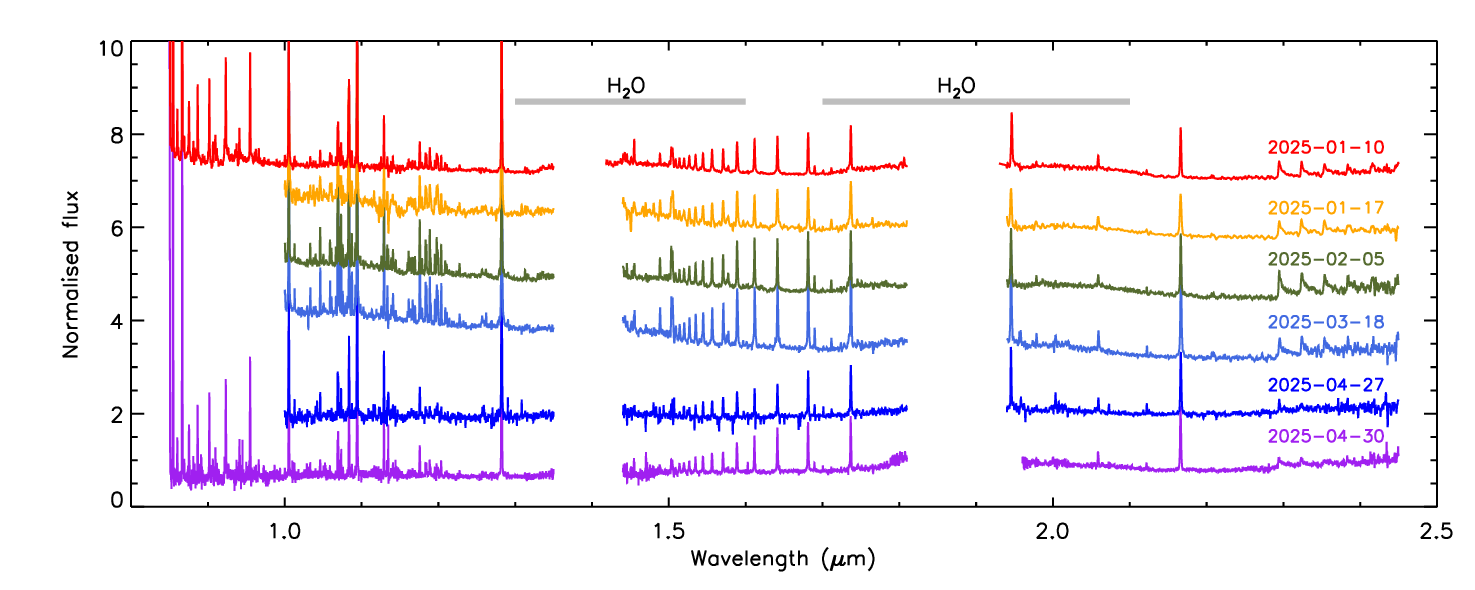}
     \caption{Near-infrared spectra of V557~Mon, observed during the outburst in 2024--2025. All spectra were normalised by the flux between 2.20 and 2.28 $\mu$m, and a constant flux was added for better illustration. Detailed spectral features will be marked on Fig.~\ref{fig:water_fit_feb}.}
\label{fig: NIR_spec}
\end{figure*}

\subsection{The mass accretion luminosity}
\label{sec: macc}

We calculated the mass accretion luminosity ($L_{\rm acc}$) of V557~Mon from the blue-band excess produced by the accretion shock. We assumed that V557~Mon is an M1-type star with a mass of 0.5~$M_\odot$ and a fixed extinction of $A_V = 1.8$~mag. The photospheric brightness was estimated using the empirical colours from \citet{Fang2017}, as $g-J$ = 4.24~mag for an M1-type young star. We used the $J$ band to estimate the photospheric brightness, as it is minimally affected by extinction, accretion, and thermal emission from the disk. With $J = 15.86$~mag from 2MASS and $A_V = 1.8$~mag, the dereddened $J$-band brightness is $m_J = 15.42$~mag, leading to $m_{g,\,\rm phot} = 19.66$~mag. The average observed and dereddened $g$-band magnitude is $m_{g,\,\rm obs} = 18.27$~mag during the quiescent stage. With the definition of excess magnitude \citep{Guo2018a}, we calculated an excess brightness of $g_{\rm ex} = 18.62$~mag. 

We estimated the $U$-band excess luminosity of V557 Mon using the empirical relation between $g_{\rm ex}$ and $U_{\rm ex}$ of M0-type CTTS EX Lupi, the prototype of EXor \citep[][]{Wang2023}. We acknowledge here that the veiling spectrum may be different for each CTTS. 
We obtained $U_{\rm ex} = 18.3\pm0.3$ and an excess luminosity of $\log\, (L_{\rm U,\,ex}/L_\odot) = -2.1\pm0.1$. We derived $L_{\rm acc} = 0.08 \pm 0.01~L_\odot$ using the $L_{\rm U,\,ex}$ -- $L_{\rm acc}$ relationship from \citet{Gullbring1998}. Considering a 0.3~mag uncertainty in $A_V$, the associated uncertainty in $L_{\rm acc}$ is $\sim$0.13~dex. The mass accretion rate was then calculated assuming magnetospheric accretion and $L_{\rm acc} = G M \dot{M}_{\rm acc}/R_{\star} (1 - R_{\star}/R_{\rm in})$. During the quiescent state, we adopted an inner disk radius $R_{\rm in} = 5R_{\star}$, commonly among Class II YSOs. Later, for the outburst, we argue that $R_{\rm in}$ may become much smaller. If $R_{\rm in}$ changes, the conversion from $L_{acc}$ to ${\rm M_\odot~yr^{-1}}$ also changes. The resulting quiescent mass accretion rate is $0.9\pm0.3\times10^{-8}~{\rm M_\odot~yr^{-1}}$, which is typical for M1 type Class~II YSOs \citep[e.g.][]{Alcala2017, Manara2023}. Alternatively, we calculated the $L_{acc}$ from the H~$\alpha$ line luminosity, applying the empirical relationship from \citet{Alcala2017}. We derived $L_{acc} =  0.13~L_\odot$ with an intrinsic scatter of 0.3 dex on the empirical relationship. This is consistent with the calculation based on blue-band excess.

We derived the quiescent luminosity of V557~Mon using the bolometric correction from \citet{Fang2017}, $m_J = 15.42$~mag and $d = 1.44$ kpc. We found the quiescent luminosity is 0.31~L$_\odot$, including $L_{\rm acc} = 0.08$~L$_\odot$. The estimation of stellar mass and age strongly relies on the pre-main-sequence evolution models. Adopting the 2 Myr age of NGC 2244 and the quiescent luminosity, we derived the mass of V557~Mon between 0.4 and 0.5~M$_\odot$ \citep{Baraffe2017}. The corresponding $T_{\rm eff}$ (3603~K -- 3763~K) is consistent with the SpT determined in Sect.~\ref{sec: quiescent} \citep[see SpT to $T_{\rm eff}$ conversions in][]{Herczeg2014}. 

\section{The 2024--2025 outburst}
\label{sec:outburst}
\subsection{Optical light curves}

An optical outburst of V557~Mon was reported in 2024 with $\Delta G = 3.5$~mag from \textit{Gaia}. The rising stage of V557~Mon extended beyond the \textit{Gaia} time-series and finally reached the peak brightness within $\sim$60~days, with $\Delta r = 4.3$~mag from ZTF. A dip is seen on the light curve immediately after V557~Mon reaches its photometric maxima, which has been observed on several other eruptive YSOs, including Gaia24ccy \citep{Singh2026}. Based on the colour track on the CMD, we suspect the dip is linked to variable extinction, as it does not follow the colour variation derived from the slab model linking to the mass accretion events. During the 100-day brightness plateau, V557~Mon displayed low-amplitude fluctuations with colour variation similar to unstable mass accretion. In late February 2025, following a secondary brightening, V557~Mon entered a gradual fading stage. A local minimum was reached in mid-August 2025, marking the end of the outburst. Compared with the pre-outburst brightness, V557 Mon is $\sim$0.5~mag fainter after the outburst in both the $g$ and $r$ bands. Such post-outburst dimming has only been seen on V1741 Sgr \citep{Kuhn2024}.  The total duration of the event is approximately 1 year, comparable to EXor outbursts \citep[e.g.][]{Cruz-Saenz2023, Singh2026}. The optical light curves of the outburst are presented in Fig.~\ref{fig: outburst_lc}.

Here, we adopted the two-step function from \citet{Lucas2024} to fit both the rising and decaying stages of the outburst as
\begin{align}
t < t_{1/2}:\,\,\,\,\,\,\,\,\, &m(t) = m_q - \frac{m_q - m_p}{1+e^{-(t-t_{1/2})/{\tau}}}, \\
t \geq t_{1/2}:\,\,\,\,\,\,\,\,\, &m(t) = m_q - (m_q - m_p)(0.5 + 0.5(t-t_{1/2})/{2\tau})
,\end{align}
where $t_{1/2}$ is the time when the brightness is enhanced by half of the amplitude and $\tau$ is a timescale parameter. The peak and quiescent magnitudes are $m_p$ and $m_q$. We also fitted the decaying function to the fading stage. As is commonly seen among eruptive young stars, the rising timescale ($2\tau_{\rm rise} = 28$~d) is much shorter than the decaying timescale ($2\tau_{\rm decay} = 70$~d). 

\begin{figure}[!t]
    \centering
    \includegraphics[width=\linewidth]{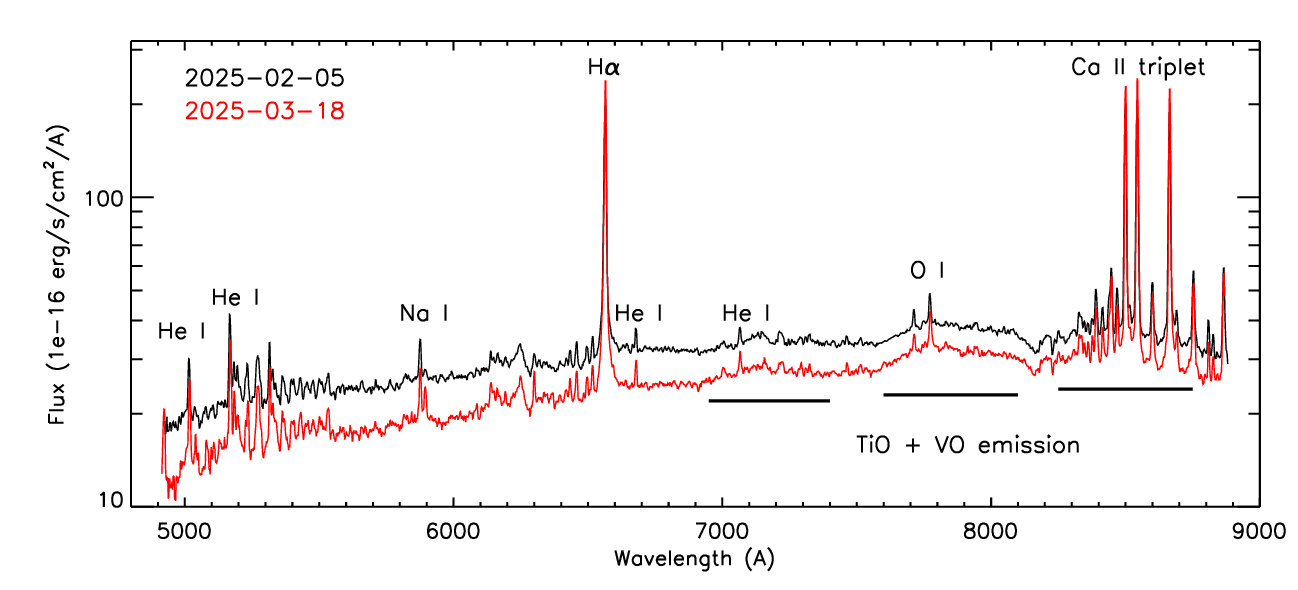}
    \caption{Two optical spectra of V557~Mon during the recent outburst. Highlighted spectral features, including TiO and VO emission bands, are marked on the plot.}
    \label{fig:opticalspectrum}
\end{figure}

\subsection{SED during the outburst}
The optical-to-NIR brightness of V557~Mon increased substantially during the outburst. The dereddened SED obtained near the outburst peak is presented in Fig.~\ref{fig:outburst_SED}, including SOAR and LCOGT $u$-band photometry. Relative to the quiescent SED, a significant flux excess is observed across the entire spectral range from the ultraviolet to the infrared. The enhanced infrared emission likely originates from a combination of a gas accretion disk, thoroughly heated by the accretion front, and a passively heated dusty disk \citep[see examples in][]{Liu2022, Das2026}.

\subsection{The outburst accretion luminosity from photometry}

Applying the method in Sect.~\ref{sec: macc} and assuming the same photospheric parameters, we measured the $L_{\rm acc}$ as a function of time, using the $g$- and $r$-band light curves (see Fig.~\ref{fig: outburst_lc}). On the optical CMD, we found that most of the data observed during the outburst lie near the slab model. However, near the photometric maxima, the extinction might be reduced by $A_V\sim0.8$~mag (see Fig.~\ref{fig: ztf_all}, around MJD 60630). Assuming $A_V = 1.8$~mag, we measured the peak $L_{\rm acc} =  4.6\, \rm L_{\odot}$, and the maximum accretion rate reached $6.3 \times 10^{-7} \rm\, \rm M_{\odot}$yr$^{-1}$, around 60 times higher than the quiescent accretion rate. Alternatively, assuming a lower $A_V = 1.0$~mag, the peak $L_{\rm acc}$ is $2.0\, \rm L_{\odot}$. The peak accretion rate is comparable to that of other EXors \citep[e.g.][]{Giannini2026}. We estimated that the total mass accreted during the outburst is of the order of $10^{-7} \rm\, \rm M_{\odot}$, $\sim$30 times the amount accreted in quiescence over the same time interval. This is reminiscent of outbursts on EX Lupi, during which the mass accretion rate is several tens to hundreds of times higher than in quiescence \citep[][]{Wang2023}. Immediately after the outburst, we found that the $g$-band brightness became fainter than the pre-outburst. A similar phenomenon has been observed among other EXors \citep[e.g. V1741 Sgr][]{Kuhn2024}, due to either enhanced line-of-sight extinction or the effect of exhausted inner disk material, resulting in a lower mass-accretion luminosity.

The innermost accretion disk of an accreting young star consists of several distinct emitting regions, including the inner gas disk, the magnetospheric free-fall region, and the stellar accretion shock. The excess luminosity of eruptive YSOs has two primary components: emission from the accretion shock and radiation from an infrared-luminous gas disk. The $L_{acc}$ calculated in Sect.~\ref{sec: macc} and above only represents the shock luminosity inferred from the blue-band excess. In \citet{Gullbring1998}, the authors assumed that radiation from the stellar surface, including the accretion shock, dominates over that from the inner accretion disk, which is reasonable for classical T~Tauri stars. However, during an accretion burst, when the gas disk extends much closer to the stellar surface (e.g. $R_{\rm in} < 5 R_*$), the free-fall velocity decreases with decreasing $R_{\rm in}$ \citep[see discussion in Sect.~\ref{sec: innerdisk} and][]{Das2026}. When $R_{\rm in}$ moves inwards, the partition of accretion energy between disk dissipation and shock emission changes. {From the change in brightness across the optical-to-NIR SED before and during the outburst of V557~Mon, we found an excess luminosity of 2.7$\sim$L$_\odot$, composed of the accretion shock luminosity and a bright accretion disk. According to the time-dependent shock luminosity around MJD 60680, we estimated the disk luminosity is 1.5 L$_\odot$, which is comparable to the accretion shock luminosity. }

\subsection{Spectra during the outburst}

The optical spectra during the outburst show typical features of EXor outbursts, such as strong H~$\alpha$ and Ca {\sc ii} triplet emission (see Fig.~\ref{fig:opticalspectrum}). We measured the $L_{\rm acc} \sim 2.0 \, \rm \rm L_{\odot}$ using the empirical relationship between H~$\alpha$ luminosity and $L_{\rm acc}$ \citep{Alcala2017} with a spread of 0.4 dex. We discovered TiO and VO emission bands, which were previously observed on a few actively accreting YSOs along with NIR CO bandhead emission \citep[e.g. V2492 Cyg, VV CrA and  LkHa225-S][]{Hillenbrand2012, Herczeg2014, Hillenbrand2022b}. The TiO emission is in contrast to the absorption bands observed from the stellar photosphere during the quiescent state. These molecules were thought to be heated by accretion-related radiation or by shocks in the disk or outflow. No photospheric absorption bands were seen on the eruptive spectra. 

The NIR spectra during the 2025 outburst are presented in Fig.~\ref{fig: NIR_spec}. These spectra show abundant emission features, including H~{\sc i} Brackett and Paschen series, He {\sc i} lines (e.g. at 1.083~$\mu$m), and various metallic transitions. We detect strong molecular emission bands, including the CO overtone bandheads and water vapour between the J/H and H/K bandpasses (see zoomed-in spectra in the appendix). The strong H~{\sc i} emission line series on V557~Mon indicates a hot and dense accretion front, suggesting the magnetospheric truncation radius of the inner gas disk approached the stellar boundary layer. The NIR spectral feature resembles other young stars undergoing high mass-accretion phases \citep[][]{Sicilia-Aguilar2023, Armeni2024, Singh2026}, where the inner disk radius becomes much smaller during an outburst than in quiescence (e.g. $R_{\rm in, outburst} < 5 R_*$). The inner gas disk becomes strongly heated during the outburst, as indicated by the NIR excess peaking bluer than 1~$\mu$m from January to March~2025.  Using the Pa~$\beta$ and Br~$\gamma$ line luminosities, we calculated the $L_{\rm acc}$ as 0.8 to 1.9~$L_\odot$ based on the empirical relationships from \citet{Alcala2017}, which is consistent with the $L_{\rm acc}$ measured from H~$\alpha$.

\begin{figure}
    \includegraphics[width=0.9\linewidth]{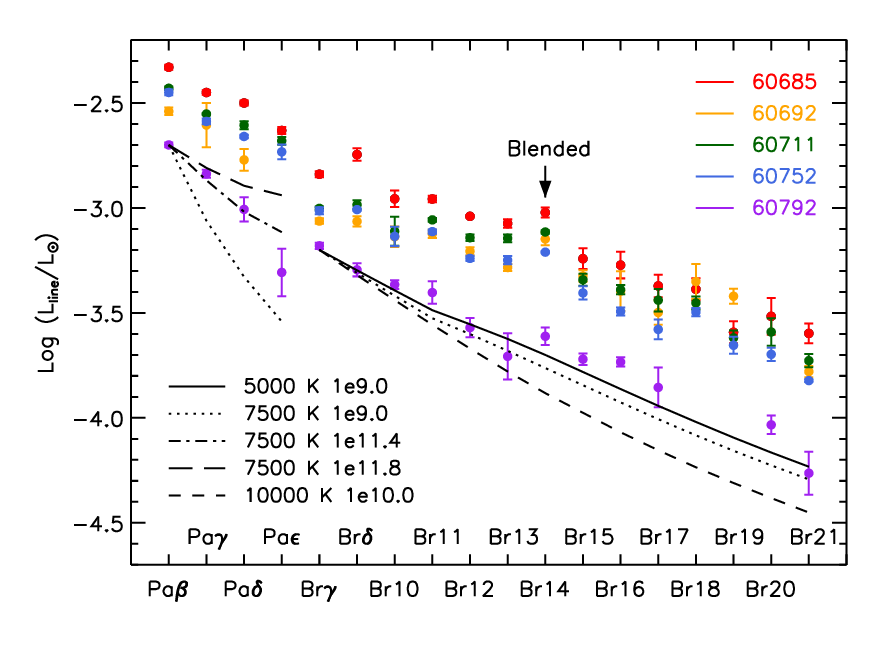}
    \caption{Luminosity of NIR H~{\sc i} emission line series on V557~Mon, from five spectroscopic epochs and dereddened by $A_V = 1.8$~mag. The MJD of each epoch is listed on the plot. Decrement models with a combination of temperature and density (cm$^{-3}$) are presented.}
    \label{fig:emissionlines}
\end{figure}

The flux ratios among different H~{\sc i} transitions provide constraints on the excitation conditions. 
The relative strengths of these lines depend primarily on the gas temperature and electron density. 
Fig.~\ref{fig:emissionlines} presents the measured H~{\sc i} line luminosities across five epochs. For each spectrum, we applied a constant extinction correction and a 1.44~kpc distance. The observed flux ratio of Pa~$\beta$ to Br~$\gamma$ is $\sim$3. In Fig.~\ref{fig:emissionlines}, we show decrement models for the Paschen series \citep[][optically thick]{Kwan2011} and for the Brackett series \citep[][optically thin]{Hummer1987}. Despite variations in line luminosity, the flux ratios among H\,{\sc i} lines remain relatively consistent across the five epochs. The best-fit models correspond to optically thick $T = 7500$~K and $n = 10^{11.4}~{\rm cm^{-3}}$ for the Paschen series, and $T = 5000$~K and optically thin $n = 10^{9}~{\rm cm^{-3}}$ for the Brackett series. However, a satisfactory fit is not achieved during the photometric maxima. {We conclude that during the accretion outburst, the near-infrared hydrogen lines should rise from a dense condition (i.e. accretion shock) where the collisional processes dominate H~{\sc i} populations. Therefore, they do not agree with the classical Case B scenario where only Ly~$\alpha$ is optically thick.} The luminosity of individual H~{\sc i} lines is listed in the appendix.

\section{Discussion}
\subsection{Molecular emission features on V557~Mon}
\label{sec: mole}

To quantify the molecular emission bands observed during the outburst, we generated synthetic spectra using slab models. We used PyExoCross \citep{Zhang2024}, a Python adaptation of the ExoCross Fortran-based code \citep{Yurchenko2018}, to create molecular emission spectra with a given local thermodynamic equilibrium temperature ($T$), surface area ($A$), and column density ($N$). PyExoCross uses the molecular transition probabilities, partition functions, and states from the ExoMol database \citep{Tennyson2016}. We generated wavelength-dependent cross-sections of four molecular species, TiO, VO, H$_2$O, and CO, without considering different isotopes. The wavelength-dependent intensity ($I_\lambda$) is defined following the methods from \citet{Salyk2022}:

\begin{equation} 
    I_\lambda = S B_\lambda(T)*(1-e^{-\tau_\lambda(T)}),
\end{equation}
where $S=A/d^2$ is the surface area of the emission region, $A$ is the emitting area, $d$ is the distance to the object, $B_\lambda(T)$ is the temperature-dependent blackbody emission of the emitting source, and $\tau_\lambda(T)=N*\sigma_\lambda(P,T)$ is the optical depth, which depends on the column density and absorption cross-section, $\sigma_\lambda$. The modelled $\sigma_\lambda$ is a function of $T$ and pressure $P$, where the latter is fixed at 1 bar for simplicity. In this work, we focus on searching for the best-fitting $T$ and $N$ values.

A 2D grid of (T, N) combinations is modelled to generate the $\sigma_\lambda$ and the model spectra. The temperature ranges from 1700 to 3600~K in steps of 100~K, whereas the range of N is from $1\times10^{20}$ to $4\times10^{20}$~cm$^{-2}$ in steps of $0.5\times10^{20}$~cm$^{-2}$. Due to the complexity of the NIR continuum spectrum, we normalised the observed spectra at 1.65 $\mu$m and then removed the continuum by fitting a third-order polynomial function. Then, we performed 2D $\chi^2$ fitting in T and N space. Finally, we identified the best-fit parameters as the mean of the best ten combinations with the minimum $\chi^2$ values. Some example model spectra are presented in the appendix.

\begin{figure*}
  \centering
     \includegraphics[height=6cm]{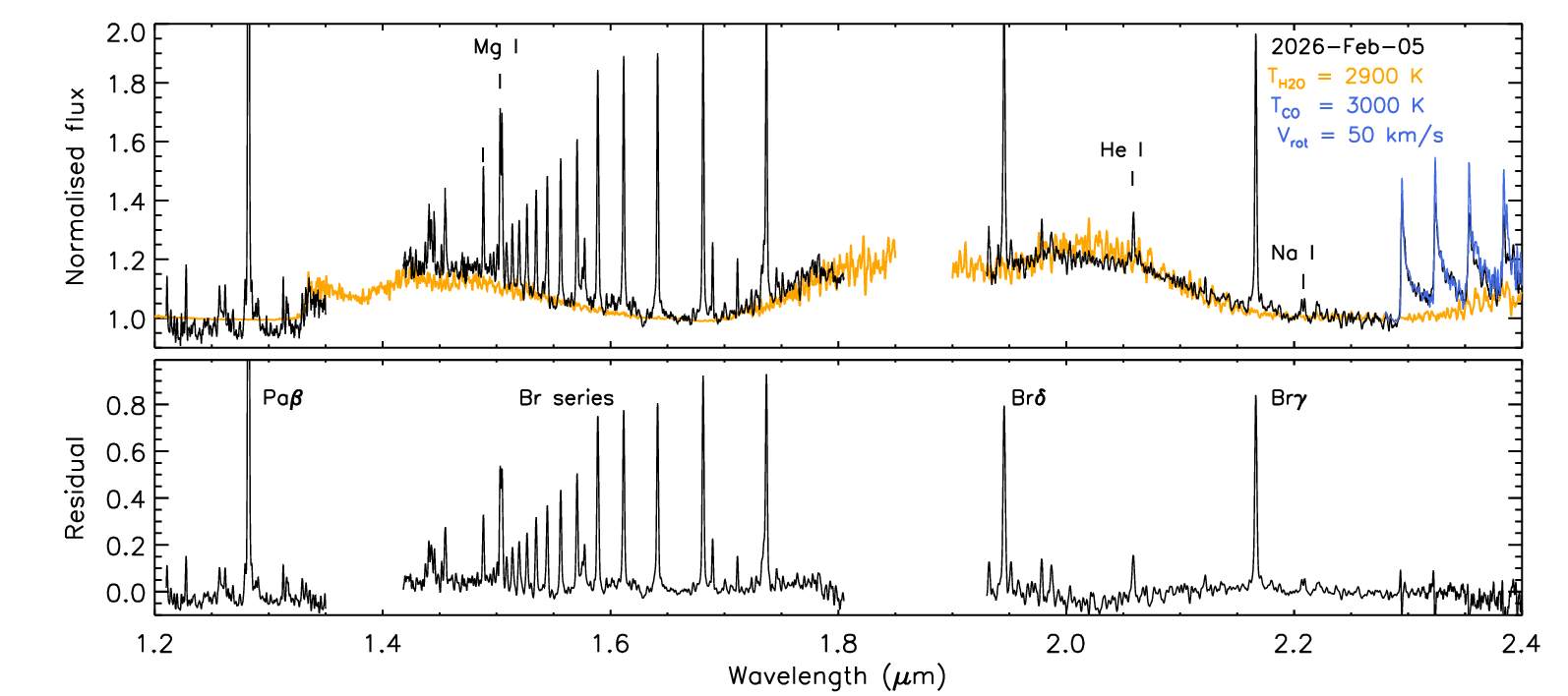}
     \caption{\textit{Upper:} Normalised NIR spectrum of V557~Mon observed on February 5 2025, during the outburst. A third-order polynomial fit removes the continuum emission. A water-emission model is shown as the orange line, and a CO model is presented as the blue line. The best-fit parameters are shown in the legend. \textit{Lower:} Fitting residual. Some emission lines are labelled on the plot. }
\label{fig:water_fit_feb}
\end{figure*}

\subsection{CO band emission}

The CO $\Delta\nu = 2$ rovibrational transitions are prominently detected between 2.29 and 2.45~$\mu$m, a feature commonly seen in low-mass stars (in absorption) and in accreting YSOs \citep[in emission,][]{Carr1989, Najita2003, Lorenzetti2012}. The presence of CO bandhead emission is widely interpreted as evidence for a hot gaseous accretion disk \citep{Connelley2010}, sometimes exhibiting variability on day-to-day timescales \citep{Guo2020}. Clear CO bandhead emission is observed in all epochs except one during the outburst. The spectrum taken on April 27 has an insufficient S/N and is excluded from the following analysis.

Following the method of \citet{Contreras2017b}, we first removed the continuum contribution (including the water vapour emission band) using a linear function. We then modelled the CO bandhead emission on the normalised spectra using transition probabilities and partition functions from \citet{Li2015}, varying parameters such as radial velocity, column density, and temperature. A Keplerian rotational broadening of 50~km~s$^{-1}$ was introduced to the synthetic spectra by a convolution function, improving the agreement with the model of the bandheads, which are better observed from the FIRE spectra. A positive correlation is found between the overall NIR flux and the derived CO bandhead temperature. As the accretion rate increases during the outburst, the inner disk radius moves inwards and the viscous heating rises, raising the gas temperature approaching the photospheric temperature. A hotter disk also emits more NIR continuum flux, producing the observed positive trend.

\begin{table*}[!h]
\centering
\caption{Best-fit parameters of molecular emission bands.}
\renewcommand\arraystretch{1.1}
\begin{tabular}{c | c c c c c c c c c}
\hline
\hline
Epoch  & $T_{\rm CO}$  & $N_{\rm CO}$ & $F_{\rm CO}$ & $T_{\rm H_2O}$ & $N_{\rm H_2O}$ & $F_{\rm H_2O}$ & m$_g$ \\
\hline
MJD &~K  & $10^{20}$ cm$^{-2}$  &erg s$^{-1}$ cm$^{-2}$ &~K & $10^{20}$ cm$^{-2}$ & erg s$^{-1}$ cm$^{-2}$ & mag\\
\hline
60685 & 3000 & 3.6 & $6.1 \times 10^{-14}$ & 2900 & 3.5  & $9.1 \times 10^{-13}$ & 17.58\\
60692 & 2900 & 3.4 & $3.5 \times 10^{-14}$ &  3300 & 3.2  & $4.8 \times 10^{-13}$& 17.62\\
60711 & 3100 & 6.7 & $6.9 \times 10^{-14}$ &  3100 & 3.3  & $6.7 \times 10^{-13}$& 17.29\\
60752 & 2700 & 4.6 & $3.3 \times 10^{-14}$ &  2000 & 2.3  & $3.0 \times 10^{-13}$& 17.90\\
60792 & 2300 & 2.0 & $1.8 \times 10^{-14}$ &  2000 & 1.2  & $1.8 \times 10^{-13}$& 18.57\\
\hline
\hline
\end{tabular}
\label{tab:molecular}
\end{table*}

\subsection{NIR water vapour emission bands}
Water vapour absorption bands are common features in NIR spectra, frequently observed in late-type stars and low-mass YSOs \citep[e.g.][and Fig.~\ref{fig: qspec}]{Greene1996}. Individual water emission lines have also been detected in young stars. For example, \citet{Najita2009} reported the first detection of water emission in T~Tauri stars near 2~$\mu$m. In the mid-infrared, water emission tracing the protoplanetary disks of T~Tauri stars has been observed with \textit{Spitzer} \citep[][]{Salyk2008} and more recently with \textit{JWST} \citep[][]{Romero-Mirza2024, Banzatti2025}.

Broad NIR water emission bands have been reported in a few protostars, including the protobinary system SVS~13 \citep{Carr2004} and the highly variable Class~I protostar V2492~Cyg \citep{Aspin2011, Hillenbrand2013}. V2492~Cyg is a jet-driving source exhibiting high-amplitude, stochastic variability and has experienced at least two accretion outbursts. During its 2010 outburst, strong water emission bands were clearly detected. To date, such pronounced water emission features have not been observed in any Class II YSOs during their quiescent stage.

We identified H$_2$O emission on V557~Mon  between the J/H and H/K bandpasses, also beyond 2.3~$\mu$m, blended with the CO bandheads (see Fig.~\ref{fig: NIR_spec}). When conducting the 2D $\chi^2$ fitting of water emission bands, we prioritised the spectral range between 1.7 and 2.2~$\mu$m with emission lines masked. As shown in Fig.~\ref{fig: H20model}, the width of the emission band is sensitive to the $T_{\rm eff}$, and our fitting has a grid of 100~K in $T_{\rm eff}$. The water emission band between the $J$ and $H$-bandpasses is contaminated by the Brackett continuum emission at 1.5~$\mu$m, whilst the water emission is blended with CO bandheads beyond 2.3~$\mu$m. 

An example of the water emission model {for the spectrum taken on MJD~60711} is presented in Fig.~\ref{fig:water_fit_feb}, along with a polynomial continuum and a CO bandhead emission model. The best-fit temperatures of the CO and water molecules are $T_{\rm CO} = 3000 \,\rm K$ and $T_{\rm H_2O} = 2900 \,\rm K$. The remaining water models and other epochs are presented in the appendix. The fitting results of all five epochs are presented in Table \ref{tab:molecular}. We also present the integrated flux of CO and H$_2$O models in Table \ref{tab:molecular}.  We concluded a positive relationship between the effective temperature of different molecules and the overall stellar brightness. The 3000~K temperature of CO overtone emission is commonly seen among actively accreting or eruptive young stars \citep[see][]{Contreras2017b, Guo2020}. However, the water vapour band on V557~Mon has a much higher temperature than those detected in previous works \citep[1400 K;][]{Najita2009}. {Similar to $T_{\rm eff}$, the column density of CO and H$_2$O also shows a positive relationship with the overall flux. This correlation suggests that the outburst heated and expanded the inner gas accretion disk. During the decaying stage, this warm and dense structure subsequently fades as the system returns to quiescence.}

\subsection{Rejuvenation of an accretion disk}

The optical and NIR spectra of V557~Mon displayed abundant molecular emission bands during the outburst, including TiO, VO, CO, and H$_2$O. Follow-up spectra after the outburst revealed no visible molecular emission features, indicating that the emission bands were transient and associated with a hot-gas disk produced during the outburst. The NIR spectra of V557~Mon closely resemble those of the actively accreting Class~I YSO V2492~Cyg.

\begin{figure*}
  \centering
     \includegraphics[height=4.5cm]{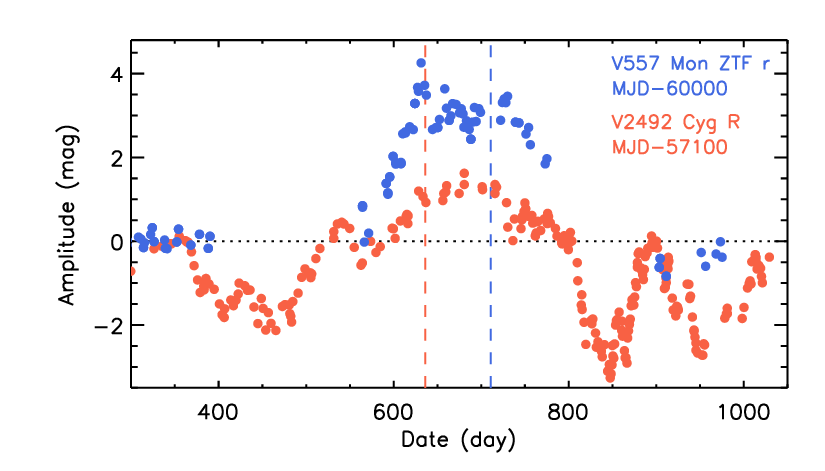}
     \includegraphics[height=4.5cm]{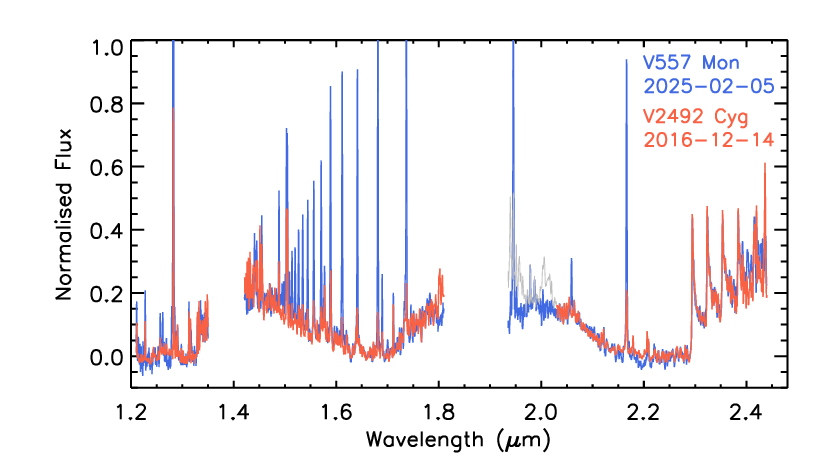}
     \caption{\textit{Left:} Optical ZTF $r$ and R-band light curves of V557~Mon and V2497 Cyg. The date of observation was aligned to illustrate the similarity of the brightening events. We subtracted the quiescent magnitudes to emphasise the photometric variation. Here, a positive amplitude means the target is brighter than the quiescent stage. The two spectroscopic epochs are shown by vertical dashed lines. \textit{Right:} Normalised NIR spectra of V557~Mon and V2497 Cyg. The V557~Mon spectrum is dereddened by $A_V = 1.8$~mag. The grey line represents a spectral range of V2492 Cyg with poor telluric correction.}
\label{fig:V557_V2492}
\end{figure*}

In Fig.~\ref{fig:V557_V2492}, we compare the optical light curves and NIR spectra of the Class~II YSO V557~Mon and the Class~I YSO V2492~Cyg (PI: Herczeg). The $R$-band light curves of V2492~Cyg were obtained from AAVSO\footnote{American Association of Variable Star Observers, \url{https://www.aavso.org/}}, \citet{Giannini2018}, and \citet{Ibryamov2018}. The NIR spectra of these two YSOs show remarkable similarity, including nearly identical water and CO emission bands, indicative of comparable physical conditions in their inner accretion disks. However, V2492 Cyg was not in a typical EXor outburst during the spectroscopic observation, and some emission lines are jet-driven. Similar emission band-dominated spectra were frequently observed on this Class I YSO \citep{Hillenbrand2013}, except for the two spectra obtained during the faintest stage (June 2011). In both stars, the molecular emission-band strength has a positive correlation with the stellar continuum brightness. The major difference between the two spectra is in the permitted atomic lines, particularly H I. The strong Brackett series observed in V557~Mon is expected, given that it was undergoing a magnetospheric accretion burst.

The presence of molecular emission bands is clearly linked to the eruptive event on V557~Mon, originating from a temporarily heated gas disk. A plausible interpretation is that the outburst represents a short-term rejuvenation of its inner accretion disk. In a quiescent low-mass Class~II system, the inner disk remains near thermal equilibrium, with the dust sublimation front and molecular gas layers largely shielded from direct stellar irradiation \citep[][]{Hartmann2016}. However, a sudden increase in the mass accretion rate by one to two orders of magnitude can deposit substantial radiative energy into the disk surface, temporarily elevating its temperature above 2000~K. Such thermal restructuring effectively replicates the physical conditions characteristic of younger, more active Class~I disks (e.g. V2492~Cyg). During the decline, as the accretion rate decreases and the disk cools, the column density of molecular species rapidly declines, or they become self-shielded within cooler layers, leading to the disappearance of the emission bands observed in the post-outburst spectra. The cooling timescale of the disk is only 150 days.

\subsection{Constraints on the outbursting inner-disk response}
\label{sec: innerdisk}
The observed rise and decay timescales of the event provide useful order-of-magnitude constraints on the spatial scale of the disk region involved in the outburst. In a standard $\alpha$ disk, the local thermal response time is $t_{\rm th}\sim(\alpha\Omega_{\rm K})^{-1}$ \citep[e.g.][]{Shakura1973}, where $\Omega_{\rm K}$ is the Keplerian angular frequency. Adopting the stellar mass inferred for V557~Mon, $M_\star \simeq 0.5\,M_\odot$, and a representative viscosity parameter of $\alpha=0.03$, the observed characteristic rise and decay timescales, $\simeq 28$ d and $\simeq 70$ d, correspond to radii of $R \approx 0.047$ AU and $R \approx 0.087$ AU. For a plausible range, $\alpha=0.01$--0.1, these values span $R \approx 0.023$--0.105 AU for the rise and $R \approx 0.042$--0.194 AU for the decay. Thus, the eruptive event most likely involved only the innermost part of the disk.

An independent estimate can be obtained from the molecular excitation temperatures. If the CO/H$_2$O-emitting layer is approximated as a radiatively heated disk surface, the equilibrium temperature can be written as
\begin{equation}
T(R)\simeq \left(\frac{L}{16\pi\sigma R^2}\right)^{1/4},
\end{equation}
where $L$ is the irradiating luminosity and $\sigma$ is the Stefan--Boltzmann constant. Using the peak luminosity scale of $L \simeq 4.6\,L_\odot$, the fitted molecular temperatures of 3000, 2500, and 2000~K (with a grid size of 100~K) imply characteristic radii of $R \approx 0.015$, 0.021, and 0.033 AU, respectively.
These values should be regarded as order-of-magnitude estimates of the characteristic emitting scale under a simple irradiation assumption.

The strong increase in the mass accretion rate implies substantial compression of the magnetosphere. Using standard magnetospheric truncation theory, $R_t \propto \dot{M}^{-2/7}$ \citep[e.g.][]{Ghosh1979}, the increase from $\dot M_{\rm quiet}\simeq 0.9\times10^{-8}\,M_\odot\,{\rm yr}^{-1}$ to $\dot M_{\rm burst}\simeq 6.3\times10^{-7}\,M_\odot\,{\rm yr}^{-1}$ reduces the truncation radius by a factor of $\sim0.3$.

The quiescent truncation radius can be estimated using the standard CTTS truncation-radius scaling \citep{Hartmann2016},
\begin{equation}
R_{\rm tr}=12.6\,B_3^{4/7}R_2^{12/7}M_{0.5}^{-1/7}\dot{M}_{-8}^{-2/7},
\end{equation}
where $B_3$ is the stellar dipolar magnetic field in units of 1 kG, $R_2$ is the stellar radius in units of $2\,R_\odot$, $M_{0.5}$ is the stellar mass in units of $0.5\,M_\odot$, and $\dot{M}_{-8}$ is the mass accretion rate in units of $10^{-8}\,M_\odot\,{\rm yr}^{-1}$.

Adopting representative stellar parameters of $B_\star \simeq 1$ kG, $M_\star \simeq 0.5\,M_\odot$, and $R_\star \simeq 2\,R_\odot$, together with the measured quiescent accretion rate of $\dot M_{\rm quiet}\simeq0.9\times10^{-8}\,M_\odot\,{\rm yr}^{-1}$, yields a quiescent truncation radius of approximately $6$--$7\,R_\star$, corresponding to $\sim0.06$--0.07 AU. Applying the same scaling to the burst accretion rate gives a truncation radius of approximately $2\,R_\star$, or $\sim0.02$ AU. This indicates that the burst likely brought the inner disk substantially closer to the star, while still remaining in a regime of active magnetospheric accretion, consistent with the strong H~$\alpha$ emission observed during the outburst \citep[see more discussion in][]{Das2026}.

At the same time, hydrostatic equilibrium in the heated inner disk implies $H/R \propto c_s/v_{\rm K} \propto \sqrt{T}$. Raising the characteristic temperature from $\sim 1500$~K in quiescence to $\sim 3000$~K during the burst increases the scale height by a factor of $\sqrt{2}\approx 1.41$. This moderate but important vertical expansion increases the solid angle subtended by the inner rim and enhances the reprocessing of accretion luminosity into the NIR, naturally supporting the observed strong continuum excess and molecular emission bands. In summary, the outburst most likely involved a reduced magnetospheric truncation radius
and the short-lived heating and vertical expansion of the innermost molecular disk, rather than a global restructuring of the broader circumstellar environment.

\section{Summary}

In this paper, we presented an EXor eruptive event that occurred on a low-mass CTTS. The outburst started in late 2024, and the entire event lasted for 1 year, with a photometric amplitude of $\Delta r \sim 4.3$~mag. We conducted photometric and spectroscopic follow-up observations under the caught-on-fire observation campaign. Our main discoveries are: 

\begin{itemize}
    \item V557~Mon is a disk-bearing Class II YSO located in the open cluster NGC2244, with an age of 2 Myr. The pre-outburst photometric analysis indicates that it is an M1-type, 0.5 $M_\odot$ star with an extinction of $A_V = 1.8\pm0.3$~mag. The quiescent accretion rate is $\dot{M}_{\rm acc} = 0.9 \pm 0.3 \times 10^{-8} M_\odot/\rm yr$.

    \item During the outburst, V557~Mon spent around 60 days reaching the photometric maximum. After staying on the brightness plateau for another 100 days, it entered a gradual decay phase. The entire event finished in about 1 year. 

    \item Using the $u$- and $g$-band brightness, we estimated the mass accretion rate of V557~Mon reached a peak accretion rate of $6.3 \times 10^{-7} M_\odot/\rm yr$ and accretion luminosity of  $4.6\, \rm \rm L_{\odot}$, typical of EXor objects \citep[see Fig.~16 and~17 in][]{Giannini2026}.

    \item Optical to NIR spectra were taken during the outburst. We observed strong molecular emission bands (VO, TiO, H$_2$O, and CO) having a positive relationship between the strength of the emission bands and the overall brightness of the star. 

    \item We developed emission models for H$_2$O and CO molecular bands using the ExoMol database. The best-fit effective temperature of the molecules ranges between 2000 and 3000 K, indicating that a hot inner accretion disk appeared during the eruptive event, possibly heated by the accretion burst.

    \item We compared the NIR spectra of V557~Mon with an active Class I YSO, V2492 Cyg, taken during a local brightening phase. The two spectra have great similarity, indicating a rejuvenation of an evolved Class II disk during an EXor outburst, which shares a similar physical property to an actively accreting Class I disk.
    
\end{itemize}

Our observations of V557~Mon demonstrate that a Class II inner accretion disk can temporarily revert to an earlier, more vigorous accretion state during an eruptive event. Such eruptive accretion events are likely to alter the chemical composition and thermal structure of the inner disk, affect dust processing, and potentially impact subsequent planet formation and evolution. Unfortunately, spectroscopic coverage during the first 100 days of the outburst (particularly the rising phase) was lacking, limiting our ability to constrain the physical trigger of the eruption. In the Rubin LSST era, incorporating machine‑learning‑based alert broker systems will be essential for rapid identification and spectroscopic follow‑up of eruptive YSOs.

\section{Data availability}
Table A.1 is only available in electronic form at the CDS via anonymous ftp to cdsarc.u-strasbg.fr (130.79.128.5) or via \url{http://cdsweb.u-strasbg.fr/cgi-bin/qcat?J/A+A/}.

\begin{acknowledgements}
We thank the anonymous referee for the careful review, which helped us improve the clarification and quality of the article.
This work is supported by the China-Chile Joint Research Fund (CCJRF No.2301) and the Chinese Academy of Sciences South America Center for Astronomy (CASSACA) Key Research Project E52H540301. ZG is supported by FONDECYT Iniciación project 11260176. JO acknowledges support from ANID Becas/Doctorado nacional/2026-21262687. ZG and CM are funded by the project ALMA-ANID 31240014. MM acknowledges financial support from FONDECYT Regular 1241818. VE acknowledges support from the Ministry of Science and Higher Education of the Russian Federation (State contract FENW-2026-0028). JB and RK thank the support from FONDECYT Regular project No. 1240249 and 1261142.

C.L. gratefully acknowledges financial support from ANID - MILENIO - NCN2024\_112.
C.C.P. was supported by the National Research Foundation of Korea (NRF) grant funded by the Korean government (MEST; No. 2019R1A6A1A10073437)"
M.W. is supported by the National Natural Science Foundation of China (grant No.124B2058)

RKY gratefully acknowledges the support from the Fundamental Fund of Thailand Science Research and Innovation (TSRI) (Confirmation No. FFB690078/0269) through the National Astronomical Research Institute of Thailand (Public Organization).

This publication makes use of data products from the Near-Earth Object Wide-field Infrared Survey Explorer (NEOWISE), which is a joint project of the Jet Propulsion Laboratory/California Institute of Technology and the University of Arizona. NEOWISE is funded by the National Aeronautics and Space Administration. This research has made use of the NASA/IPAC Infrared Science Archive, which is funded by the National Aeronautics and Space Administration and operated by the California Institute of Technology. Based on observations obtained at the Southern Astrophysical Research (SOAR) telescope, which is a joint project of the Minist\'{e}rio da Ci\^{e}ncia, Tecnologia e Inova\c{c}\~{o}es (MCTI/LNA) do Brasil, the U.S. National Science Foundation NOIRLab, the University of North Carolina at Chapel Hill (UNC), and Michigan State University (MSU). This work makes use of observations from the Las Cumbres Observatory global telescope network.

This work makes use of observations from the Las Cumbres Observatory global telescope network.

Based on observations obtained through the Astronomical Event Observatory Network (AEON), a joint endeavor of the Las Cumbres Observatory and of NSF NOIRLab, which is managed by the Association of Universities for Research in Astronomy (AURA) under a cooperative agreement with the U.S. National Science Foundation.

Based on observations collected with the Goodman/Triplespec at Cerro Pachon, Chile, under the programme allocated by the Chilean Telescope Allocation Committee (CNTAC), no CN2025A-14 and CN2025B-23.

This research is based on observations made with the Thai Robotic Telescope under program ID [TRTC12A\_013], which is operated by the National Astronomical Research Institute of Thailand (Public Organization).

The authors acknowledge the use of artificial intelligence language models (ChatGPT and Grammarly) for assistance with text editing.
\end{acknowledgements}

\bibliographystyle{aa}
\bibliography{reference}
\begin{appendix}

\section{Photometric data}
Here, we present a machine-readable table of the multi-wavelength photometric light curves of V557~Mon. The data obtained from online catalogues (e.g. \textit{Gaia}, ZTF, ATLAS, etc) are marked with "c", and our PI data (SOAR, LCOGT, SWOPE, LOT and TRT) are labelled as "p". In Fig. \ref{fig:quiescent_others}, we present the optical-to-infrared light curve during the quiescent stage. A short-timescale burst was observed by ZTF, \textit{Gaia} and \textit{NEOWISE} time-series in 2021, with $\Delta r = 2$~mag, $\Delta W1 = 0.7$~mag and $\Delta W2 = 0.8$~mag.

\begin{figure}[!b]
  \centering
     \includegraphics[height=6cm]{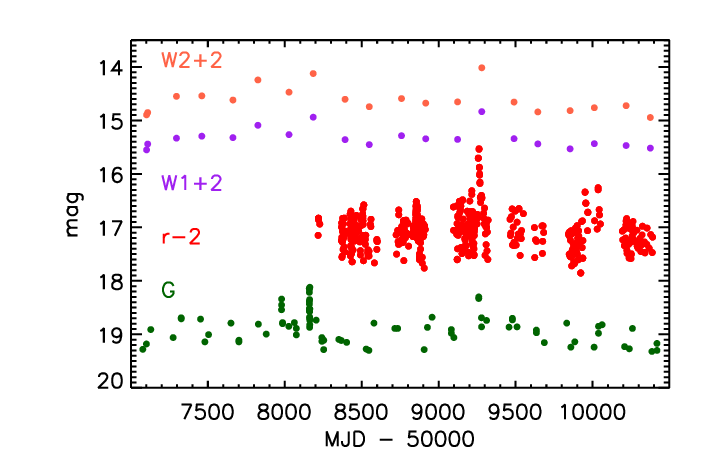}
     \caption{Quiescent light curves of V557~Mon obtained from ZTF, \textit{Gaia} and \textit{NEOWISE} surveys. The short-timescale low-amplitude burst in 2021 was captured by ZTF and \textit{NEOWISE} photometry.}
\label{fig:quiescent_others}
\end{figure}

\begin{table}[!b]
    \centering
    \caption{Photometric data of V557~Mon}
    \renewcommand\arraystretch{1.2}
    \begin{tabular}{cccccc}
    \hline
    \hline
        Telescope & MJD & band &~mag & error & mark\\
    \hline
     2MASS & 51496.35 & $J$ & 15.86 & 0.08 & c \\
     2MASS & 51496.35 & $H$ & 14.94 & 0.07 & c \\
     2MASS & 51496.35 & $K_s$ & 14.29 & 0.07 & c \\
     Pan-STARRS    & 56141.46 & $g$ & 20.59 & 0.04 & c\\
    Pan-STARRS    & 56141.46 & $r$ & 19.41 & 0.09 & c\\
    Pan-STARRS    & 56141.46 & $i$ & 18.36 & 0.01 & c\\
    Pan-STARRS    & 56141.46 & $z$ & 17.74 & 0.01 & c\\
    \textit{Gaia}   & 57077.07 & $G$ & 19.28 & - & c\\
     SOAR    & 60988.30 & $g$ & 20.85 & 0.20 & p \\
     SWOPE    & 61039.20 & $u$ & 21.63 & 0.45 & p \\
     SWOPE    & 61039.22 & $g$ & 20.53 & 0.19 & p \\
     SWOPE    & 61039.23 & $r$ & 19.41 & 0.34 & p \\
     SWOPE    & 61039.23  & $i$ & 18.32 & 0.14 & p \\
     LCOGT & 60670.24 & $u$ & 18.31 & 0.07& p \\
    LCOGT & 60669.63 & $g$ & 17.57 & 0.03& p \\
    LCOGT & 60669.64 & $r$ & 16.52 & 0.02 & p \\
ZTF & 58216.18 & $r$ & 19.15 & 0.05 & c \\
ZTF & 58367.48 & $g$ & 20.76 & 0.18 & c \\
ATLAS &  57670.64 & $o$ & 18.91 & 0.27& r \\
     ... & ... & ... & ... & ... & ... \\
    \hline
    \hline
    \end{tabular}
    \vspace{5pt}
    \tablefoot{ A full machine-readable version is presented in the online supplementary material. Data obtained from online catalogues are marked as "c", and our own PI photometric data are marked as "p".}
    \label{tab:placeholder}
\end{table}

\section{Quiescent spectra}
Here, we present two examples of fitting spectral templates to the quiescent optical spectrum and photometry of V557~Mon. The spectral templates are from \citet{Claes2024}. 
\begin{figure}[!h]
  \centering
     \includegraphics[height=6cm]{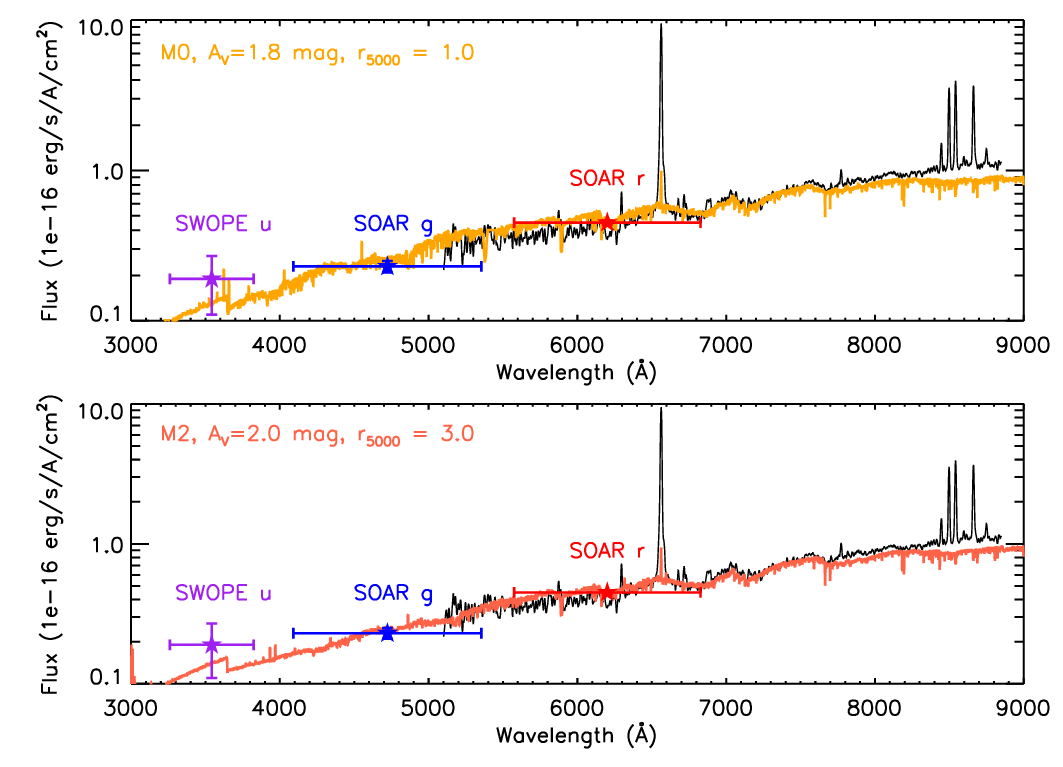}
     \caption{We present spectroscopic templates \citep{Claes2024} with the quiescent spectra of V557~Mon. We added a blue-band excess from a slab model to the photospheric templates (M0 and M2), then reddened the spectral models.}
\label{fig:quiescent_others}
\end{figure}

\section{Molecular models from ExoMol}
\begin{figure}[!h]
    \centering
    \includegraphics[width=0.9\linewidth]{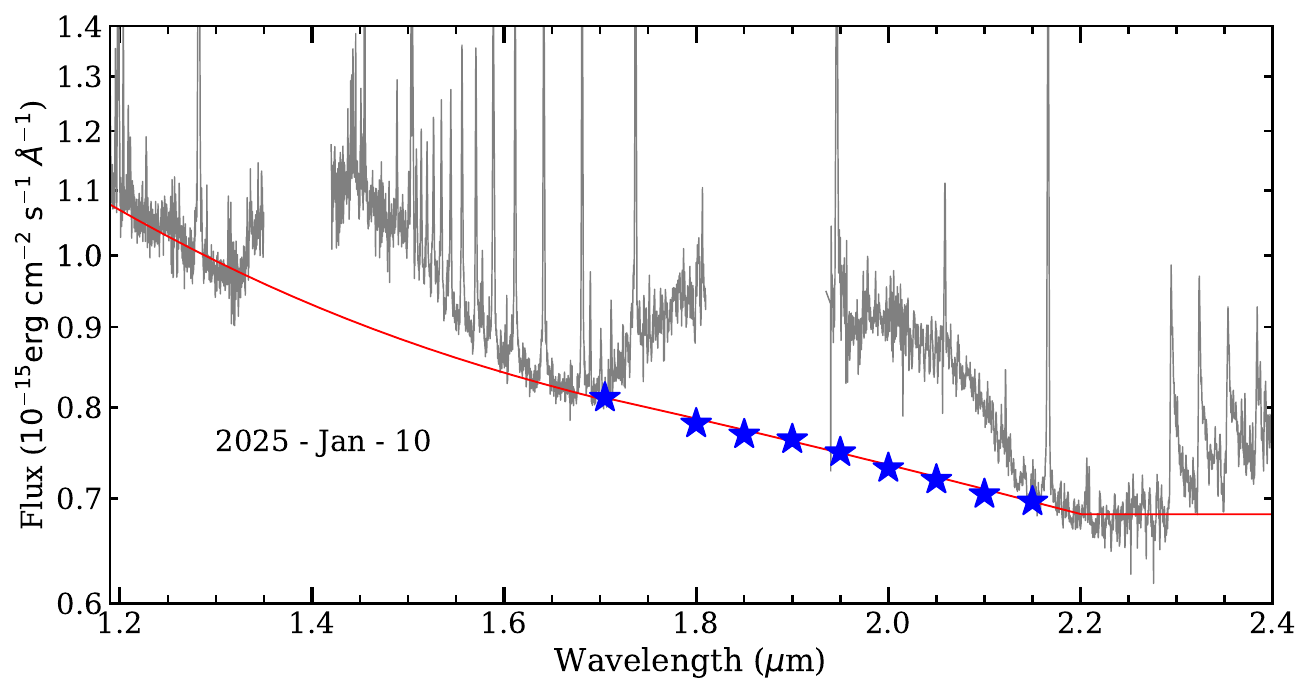}
    \caption{NIR spectrum of V557~Mon observed on MJD 60685. We extract the continuum using a 3rd-order polynomial function. The blue stars display the artificial points used to control the continuum around the 1.8 $\mu$m water emission band region.}
    \label{fig:continuumwithwater}
\end{figure}
\begin{figure}
  \centering
     \includegraphics[height=5.3cm]{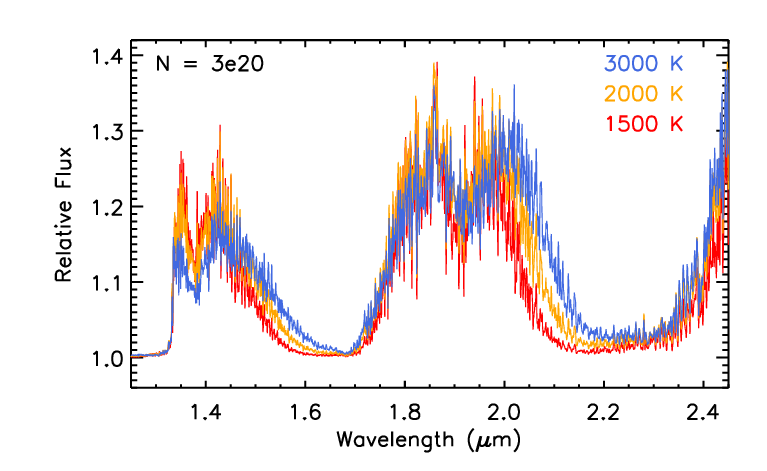}
     \includegraphics[height=5.3cm]{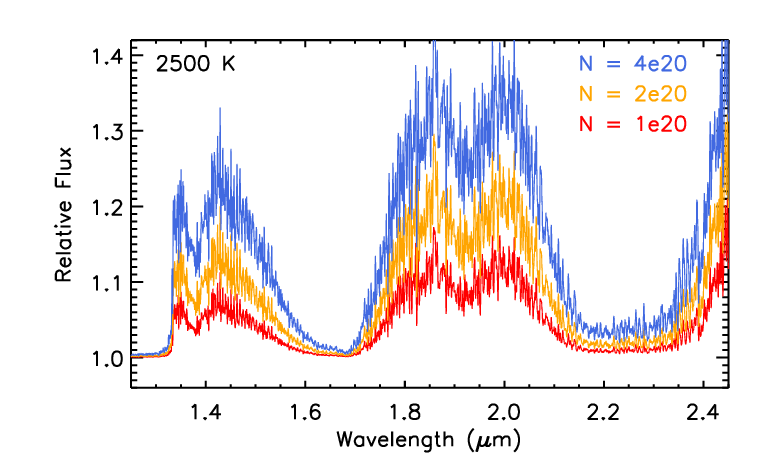}     
     \caption{Examples of the water vapour slab models we generated from the ExoMol molecular database. \textit{Upper}: models with fixed column density of $3\times10^{20}\rm cm^{-2}$ and temperature ranging between 1500 to 3000 K. \textit{Lower}: models with fixed temperature of 2500~K and column density ranging between $1\times10^{20}\rm cm^{-2}$ and $4\times10^{20}\rm cm^{-2}$.}
\label{fig: H20model}
\end{figure}

{To study the molecular emission bands in the near-infrared spectra, we first fit the continuum flux between 1.2 and 2.45 $\mu$m using a third-order polynomial function. We added a few artificial control points in the gap between $H$ and $K$-bandpasses to avoid introducing extra curvature to the water emission band. Our continuum fitting results are presented in Fig.~\ref{fig:continuumwithwater}. 
}

We utilised the ExoMol database to generate slab models of molecular emission bands, including water vapour, CO, and TiO. Here, we considered only optically thin cases, as the emission bands originate from a thin layer of hot gas in the inner disk. In Fig.~\ref{fig: H20model}, we present the water vapour slab models with effective temperature ranging between 1500 and 3000~K and column density between $1\times10^{20}\rm cm^{-2}$ and $4\times10^{20}\rm cm^{-2}$. Notably, at a fixed column density, the emission bandwidth at 2.0 - 2.1 $\mu$m is most sensitive to the effective temperature. On the other hand, with a fixed temperature, the height of the emission band is sensitive to the column density. 

\begin{figure}
  \centering
     \includegraphics[width=9cm]{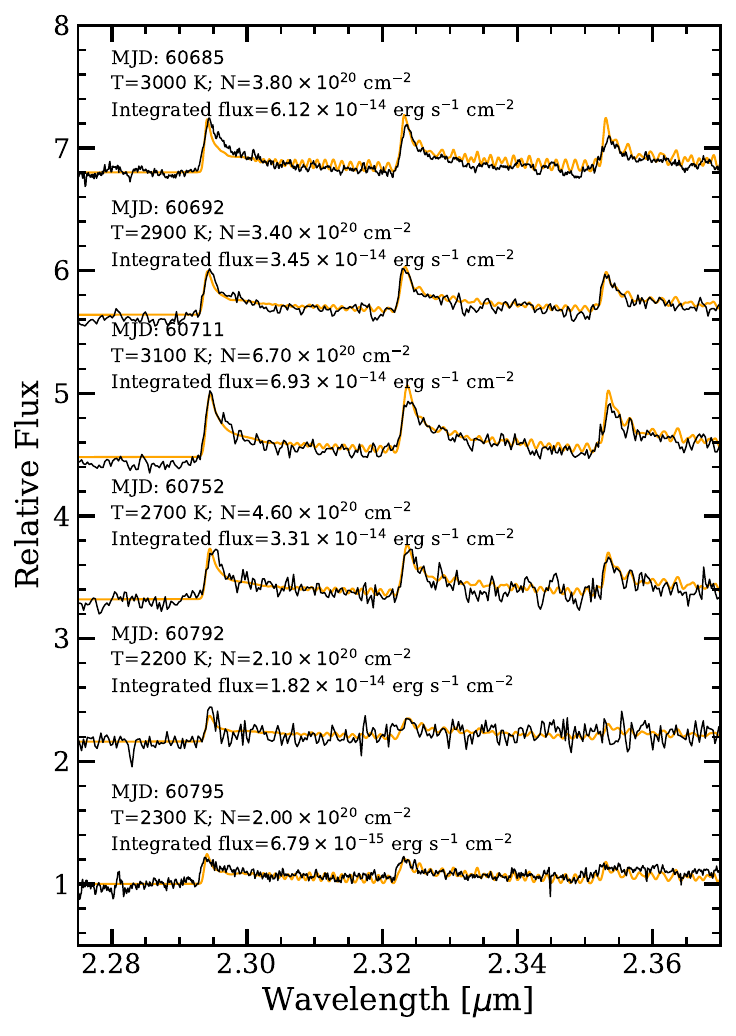}
     \caption{CO bandhead models of six spectral epochs. The best-fit parameters and integrated flux of the models are shown as legends. The observation dates in MJD are listed in the legend.}
\label{fig: comodel}
\end{figure}

For V557~Mon, the strong Brackett line series and Brackett continuum in the $H$-bandpass spectrum led us to omit a water emission model between 1.4 and 1.6 $\mu$m. We first normalised the extinction-corrected spectrum to the flux at approximately 1.7 $\mu$m. Then, we removed the stellar continuum by applying polynomial functions. We used the relative height of the 1.8 $\mu$m emission feature to find the best-fit column density. Finally, we fit the effective temperature using the slope between 2.0 and 2.2 $\mu$m. When fitting the broadband water vapour emission, we did not account for Keplerian broadening due to the spin of the hot gas disk. The best-fit model for each spectral epoch is presented below. We also fit the CO emission bands using the ExoMol models. The results of each epoch are presented in Fig.~\ref{fig: comodel}.

In addition, we compare the outburst spectrum from V557 Mon with another EXor event observed by our group, using the same instrument setup on TSpec (Singh et al., in prep). V557 Mon exhibits stronger emission lines and molecular emission bands than Gaia24ccy. 
\begin{figure}
  \centering
     \includegraphics[width=9cm]{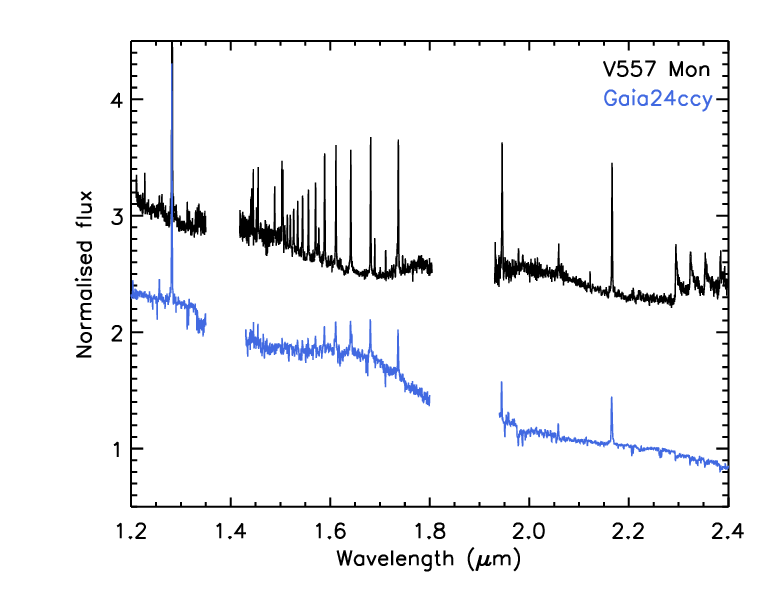}
     \caption{The outburst spectrum of V557 Mon in comparison with another EXor outburst, Gaia24ccy. Both spectra were taken by TSpec on SOAR with the same setup.}
\label{fig: compare}
\end{figure}
\begin{figure*}
  \centering
     \includegraphics[width=12cm]{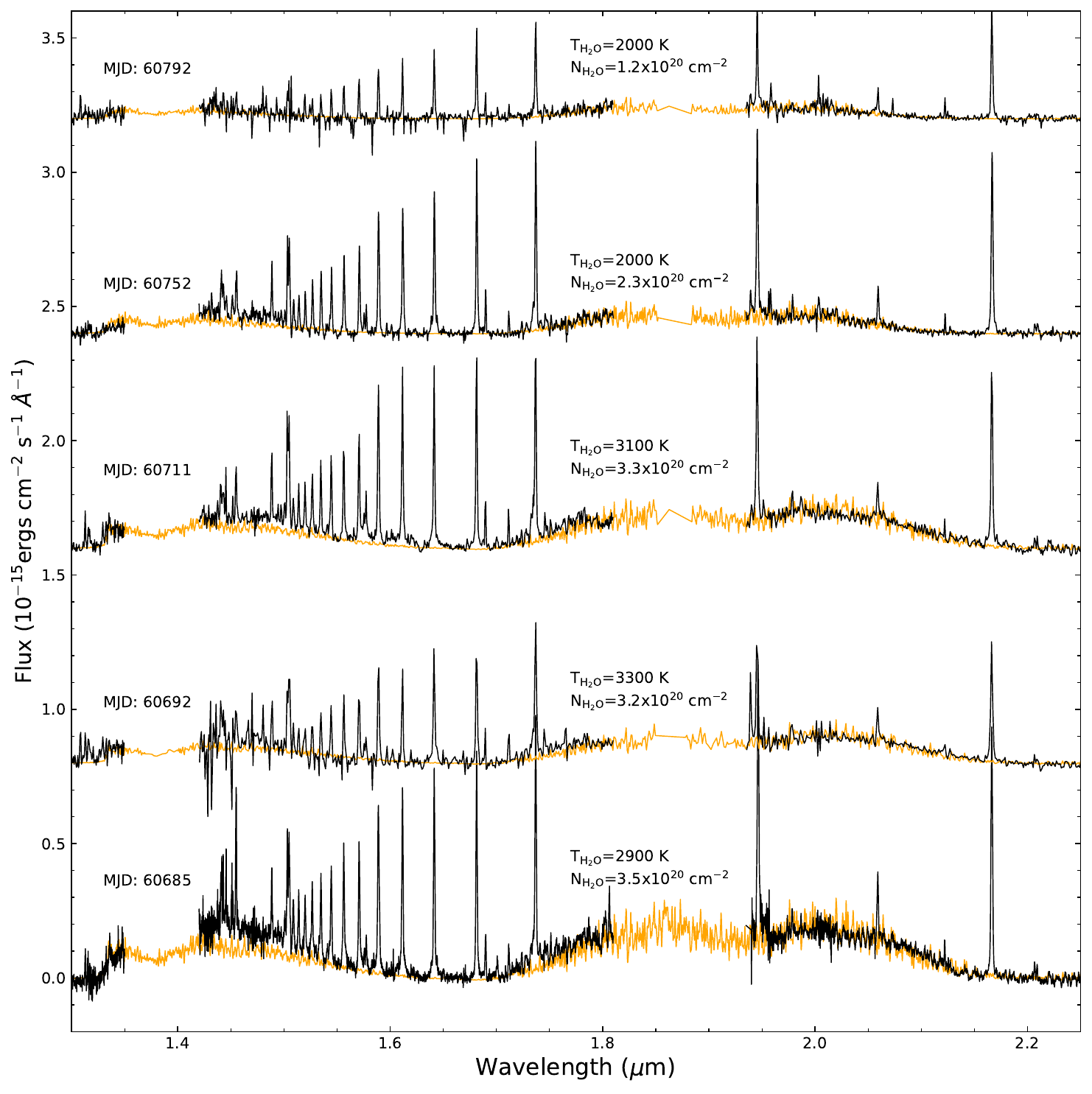}
     \caption{Hot water vapour models of six spectral epochs. The best-fit parameters and integrated flux of the models are shown as legends. The observation dates in MJD are listed in the legend.}
\label{fig: watermodels}
\end{figure*}

\begin{table*}
    \centering
     \caption{The measured flux of the Paschen and Brackett line series of V557 Mon during the outburst. }
    \renewcommand\arraystretch{1.2}
    \begin{tabular}{cccccccccccccccc}
    \hline
    \hline
    Date (MJD) & Pa~$\beta$ & Pa$\gamma$ & Pa$\delta$ & Pa$\epsilon$    \\
    \hline
    60685 & -2.33  & -2.45  & -2.50  & -2.63  \\
    60692 & -2.54  & -2.61  & -2.77  & --  \\
    60711 & -2.43  & -2.55  & -2.61  & -2.68 \\  
    60752 & -2.45  & -2.59  & -2.66  & -2.73 \\ 
    60792 & -2.70  & -2.84  & -3.01  & -3.31 \\  
     \hline
    Date (MJD) & Br~$\gamma$ & Br~$\delta$ & Br10 & Br11 & Br12 & Br13 & Br14 & Br15 & Br16  & Br17  & Br18\\
    \hline
     60685 & -2.84  & -2.75  & -2.96  & -2.96  & -3.04  & -3.07  & -3.02  & -3.24  & -3.27  & -3.37  & -3.39 \\
     60692 &-3.06  & -3.06  & -3.14  & -3.12  & -3.21  & -3.28  & -3.15  & -3.32  & -3.40  & -3.50  & -3.35 \\
     60711 & -3.00  & -2.98  & -3.11  & -3.06  & -3.14  & -3.14  & -3.11  & -3.34  & -3.39  & -3.44  & -3.45\\
      60752 & -3.01  & -3.01  & -3.14  & -3.11  & -3.24  & -3.25  & -3.21  & -3.40  & -3.49  & -3.58  & -3.50\\
      60792 & -3.18  & -3.29  & -3.37  & -3.40  & -3.57  & -3.71  & -3.61  & -3.72  & -3.73  & -3.86  & -- \\
    \hline
    \end{tabular}
    \tablefoot{The unit of the flux is $\log{(L_{\rm line}/\rm L_{\odot})}$.}
    \label{tab:placeholder}
\end{table*}
    
\end{appendix}

\end{document}